\documentclass[runningheads]{llncs}

\usepackage[pdftex]{graphicx}
	\pdfoutput=1

\usepackage{amsmath,amssymb}

\usepackage{color}
	\definecolor{color1}{rgb}{0,0,0.8}

\usepackage[ruled,linesnumbered]{algorithm2e}
  \SetKw{Break}{break}
  \SetFuncSty{textrm}
  \SetArgSty{textrm}
  \SetCommentSty{textit}
  \SetKwBlock{Forever}{repeat}{end}
  \SetKw{KwSide}{Side effects: }

\usepackage[colorlinks=true,
	linkcolor=black,
	citecolor=color1,
	filecolor=color1,
	menucolor=color1,
	urlcolor=color1,
	pdfpagelayout=TwoColumnRight,
	pdftitle={The Lazy Flipper: MAP Inference in Higher-Order Graphical Models by Depth-limited Exhaustive Search},
	pdfsubject={Optimization of Functions that Decompose According to a Graphical Model},
	pdfkeywords={Graphical Model,Markov Random Field,Conditional Random Field,MRF,CRF,Energy Minimization,Inference,MAP,Higher-Order Factors,Higher-Order Cliques,Iterated Conditional Modes,ICM,Belief Propagation,Message Passing,Dual Decomposition}]{hyperref}

\newcommand{\nmax}{n_{\textnormal{max}}}

\begin{document}
\pagestyle{headings}

\mainmatter
\title{The Lazy Flipper: MAP Inference in Higher-Order Graphical Models by Depth-limited Exhaustive Search}
\titlerunning{The Lazy Flipper}
\authorrunning{B.~Andres et al.}
\author{Bj\"{o}rn Andres, J\"{o}rg H.~Kappes, Ullrich K\"{o}the and Fred A.~Hamprecht}
\institute{HCI, IWR, University of Heidelberg\\
	\href{http://hci.iwr.uni-heidelberg.de}{http://hci.iwr.uni-heidelberg.de},
	\href{mailto:bjoern.andres@iwr.uni-heidelberg.de}{bjoern.andres@iwr.uni-heidelberg.de}
}

\maketitle
\enlargethispage{1cm}

\begin{abstract}
This article presents a new search algorithm for the NP-hard problem of 
optimizing functions of binary variables that decompose according to a 
graphical model. It can be applied to models of any order and structure. The 
main novelty is a technique to constrain the search space based on the topology 
of the model. When pursued to the full search depth, the algorithm is 
guaranteed to converge to a global optimum, passing through a series of 
monotonously improving local optima that are guaranteed to be optimal within a 
given and increasing Hamming distance. For a search depth of 1, it specializes 
to Iterated Conditional Modes. Between these extremes, a useful tradeoff 
between approximation quality and runtime is established. Experiments on 
models derived from both illustrative and real problems show that 
approximations found with limited search depth match or improve those obtained 
by state-of-the-art methods based on message passing and linear programming. 
\end{abstract}

\section{Introduction}
Energy functions that depend on thousands of binary variables and decompose 
according to a graphical model 
\cite{cowell-2007,koller-2009,lauritzen-1996,wainwright-2008} 
into potential functions that depend on subsets of all variables have been used
successfully for pattern analysis, e.g.~in the seminal works
\cite{besag-1986,boycov-2001,geman-1984,mceliece-1998}.
An important problem is the minimization of the sum of potentials, i.e.~the 
search for an assignment of zeros and ones to the variables that minimizes the 
energy. This problem can be solved efficiently by dynamic programming if the 
graph is acyclic
\cite{pearl-1988}
or its treewidth is small enough
\cite{lauritzen-1996},
and by finding a minimum s-t-cut 
\cite{boycov-2001} 
if the energy function is (permutation) submodular 
\cite{kolmogorov-2004,schlesinger-2007}. 
In general, the problem is NP-hard
\cite{kolmogorov-2004}.
For moderate problem sizes, exact optimization is sometimes tractable by means 
of Mixed Integer Linear Programming (MILP)
\cite{schrijver-1986,schrijver-2003}. 
Contrary to popular belief, some practical computer vision problems can indeed 
be solved to optimality by modern MILP solvers
(cf.~Section~\ref{section:experiments}). 
However, all such solvers are eventually overburdened when the problem size 
becomes too large. In cases where exact optimization is intractable, one has to
settle for approximations. While substantial progress has been made in this 
direction, a deterministic non-redundant search algorithm that  
constrains the search space based on the topology of the graphical model has 
not been proposed before. This article presents a depth-limited exhaustive 
search algorithm, the Lazy Flipper, that does just that. 

The Lazy Flipper starts from an arbitrary initial assignment of zeros and ones 
to the variables that can be chosen, for instance, to minimize the sum of only 
the first order potentials of the graphical model. Starting from 
this initial configuration, it searches for flips of variables that reduce the 
energy. As soon as such a flip is found, the current configuration is updated 
accordingly, i.e.~in a greedy fashion. In the beginning, only single variables 
are flipped. Once a configuration is found whose energy can no longer be 
reduced by flipping of \emph{single} variables, all those subsets of two and 
successively more variables that are connected via potentials in the graphical 
model are considered. When a subset of more than one variable is flipped, all 
smaller subsets that are affected by the flip are revisited. This allows the 
Lazy Flipper to perform an exhaustive search over all subsets of variables 
whose flip potentially reduces the energy. Two special data structures 
described in 
Section \ref{section:data-structures} 
are used to represent each subset of connected variables precisely once and to 
exclude subsets from the search whose flip cannot reduce the energy due to the 
topology of the graphical model and the history of unsuccessful flips. 
These data structures, the Lazy Flipper algorithm and an experimental 
evaluation of state-of-the-art optimization algorithms on higher-order 
graphical models are the main contributions of this article.

\enlargethispage{1mm}
Overall, the new algorithm has four favorable properties: 
(i) It is strictly convergent. While a global minimum is found when searching 
through all subgraphs (typically not tractable), approximate solutions with a 
guaranteed quality certificate 
(Section \ref{section:algorithm}) 
are found if the search space is restricted to subgraphs of a given maximum 
size. The larger the subgraphs are allowed to be, the tighter the upper bound 
on the minimum energy becomes. This allows for a favorable trade-off between 
runtime and approximation quality. 
(ii) Unlike in brute force search, the runtime of lazy flipping depends on the
topology of the graphical model. It is exponential in the worst case but can be
shorter compared to brute force search by an amount that is exponential in the 
number of variables. It is approximately linear in the size of the model for a
fixed maximum search depth.
(iii) The Lazy Flipper can be applied to graphical models of any order and 
topology, including but not limited to the more standard grid graphs. Directed 
Bayesian Networks and undirected Markov Random Fields are processed in the 
exact same manner; they are converted to factor graph models 
\cite{kschischang-2001} 
before lazy flipping. 
(iv) Only trivial operations are performed on the graphical model, namely graph 
traversal and evaluations of potential functions. These operations are cheap
compared, for instance, to the summation and minimization of potential functions
performed by message passing algorithms, and require only an implicit 
specification of potential functions in terms of program code that computes the 
function value for any given assignment of values to the variables.

Experiments on simulated and real-world problems, submodular and non-submodular 
functions, grids and irregular graphs
(Section \ref{section:experiments}) 
assess the quality of Lazy Flipper approximations, their convergence as well as 
the dependence of the runtime of the algorithm on the size of the model and the 
search depth. The results are put into perspective by a comparison with 
Iterated Conditional Modes (ICM)
\cite{besag-1986}, 
Belief Propagation (BP)
\cite{pearl-1988,kschischang-2001}, 
Tree-reweighted BP
\cite{wainwright-2005,wainwright-2008}
and a Dual Decomposition ansatz using sub-gradient descent methods
\cite{komodakis-2010,kappes-2010}.

\section{Related Work}

The Lazy Flipper is related in at least four ways to existing work. 
First of all, it generalizes Iterated Conditional Modes (ICM) for binary 
variables
\cite{besag-1986}. 
While ICM leaves all variables except one fixed in each step, the Lazy Flipper 
can optimize over larger (for small models: all) connected subgraphs of a 
graphical model. Furthermore, it extends Block-ICM 
\cite{frey-2005} 
that optimizes over specific subsets of variables in grid graphs to irregular 
and higher-order graphical models. 

Naive attempts to generalize ICM and Block-ICM to optimize over subgraphs of 
size $k$ would consider all sequences of $k$ connected variables and ignore the
fact that many of these sequences represent the same set. This causes 
substantial problems because the redundancy is large, as we show in 
Section~\ref{section:data-structures}. 
The Lazy Flipper avoids this redundancy, at the cost of storing one unique 
representative for each subset. Compared to randomized algorithms that sample
from the set of subgraphs
\cite{jung-2009,swendsen-1987,wolff-1989},
this is a memory intensive approach. Up to 8~GB of RAM are required for the 
optimizations shown in 
Section~\ref{section:experiments}.
Now that servers with much larger RAM are available, it has become a practical 
option.

Second, the Lazy Flipper is a deterministic alternative to the randomized 
search for tighter bounds proposed and analyzed in 2009 by 
Jung et al.~\cite{jung-2009}. 
Exactly as in 
\cite{jung-2009}, 
sets of variables that are connected via potentials in the graphical model are
considered and variables flipped if these flips lead to a smaller upper bound on 
the sum of potentials. In contrast to 
\cite{jung-2009}, 
unique representatives of these sets are visited in a deterministic order. 
Both algorithms maintain a current best assignment of values to the variables and
are thus related with the Swendsen-Wang algorithm 
\cite{swendsen-1987,barbu-2003} 
and Wolff algorithm 
\cite{wolff-1989}.

Third, lazy flipping with a limited search depth as a means of approximate 
optimization competes with message passing algorithms \cite{kschischang-2001,minka-2001,globerson-2007,wainwright-2008}
and with algorithms based on convex programming relaxations of the 
optimization problem
\cite{globerson-2007,werner-2007,kohli-2008,kumar-2009},
in particular with Tree-reweighted Belief Propagation (TRBP) 
\cite{wainwright-2005,wainwright-2008,kolmogorov-2006} 
and sub-gradient descent
\cite{komodakis-2010,kappes-2010}. 

Fourth, the Lazy Flipper guarantees that the best approximation found with a 
search depth $\nmax$ is optimal within a Hamming distance $\nmax$. A similar 
guarantee known as the Single Loop Tree (SLT) neighborhood 
\cite{weiss-2001}
is given by BP in case of convergence. The SLT condition states that in any 
alteration of an assignment of values to the variables that leads to a lower 
energy, the altered variables form a subgraph in the graphical model that has 
at least two loops. The fact that Hamming optimality and SLT optimality differ 
can be exploited in practice. We show in one experiment in
Section~\ref{section:experiments}
that BP approximations can be further improved by means of lazy flipping.

\section{The Lazy Flipper Data Structures}
\label{section:data-structures}

Two special data structures are crucial to the Lazy Flipper. The first data 
structure that we call a \emph{connected subgraph tree (CS-tree)} ensures that 
only \emph{connected} subsets of variables are considered, i.e.~sets of 
variables which are connected via potentials in the graphical model. Moreover,
it ensures that every such subset is represented precisely once (and not 
repeatedly) by an ordered sequence of its variables, 
cf.~\cite{moerkotte-2006}. 
The rationale behind this concept is the following: If the flip of one 
variable and the flip of another variable not connected to the first one do not
reduce the energy then it is pointless to try a simultaneous flip of both 
variables because the (energy increasing) contributions from both flips would 
sum up. Furthermore, if the flip of a disconnected set of variables reduces the 
energy then the same and possibly better reductions can be obtained by flipping
connected subsets of this set consecutively, in any order. All disconnected 
subsets of variables can therefore be excluded from the search if the connected 
subsets are searched ordered by their size. 

Finding a unique representative for each connected subset of variables is 
important. The alternative would be to consider all sequences of pairwise 
distinct variables in which each variable is connected to at least one of its 
predecessors and to ignore the fact that many of these sequences represent the 
same set. Sampling algorithms that select and grow connected subsets in a 
randomized fashion do exactly this. However, the redundancy is large. As an 
example, consider a connected subset of six variables of a 2-dimensional grid 
graph as depicted in 
Fig.~\ref{figure:cs-tree}a. 
Although there is only one connected set that contains all six variables, $208$ 
out of the $6! = 720$ possible sequences of these variables meet the 
requirement that each variable is connected to at least one of its 
predecessors. This 208-fold redundancy hampers the exploration of the search 
space by means of randomized algorithms; it is avoided in lazy flipping at the 
cost of storing one unique representative for every connected subgraph in the 
CS-tree.

The second data structure is a \emph{tag list} that prevents the repeated 
assessment of unsuccessful flips. The idea is the following: If some variables 
have been flipped in one iteration (and the current best configuration has been 
updated accordingly), it suffices to revisit only those sets of variables that 
are connected to at least one variable that has been flipped. All other sets of 
variables are excluded from the search because the potentials that depend on 
these variables are unaffected by the flip and have been assessed in their 
current state before.

The tag list and the connected subgraph tree are essential to the Lazy Flipper 
and are described in the following sections, 
\ref{section:cs-tree} 
and 
\ref{section:tag-lists}. 
For a quick overview, the reader can however skip these sections, take for 
granted that it is possible to efficiently enumerate all connected subgraphs of
a graphical model, ordered by their size, and refer directly to the main 
algorithm 
(Section \ref{section:algorithm} 
and 
Alg.~\ref{algorithm:lazy-flipper}). 
All non-trivial sub-functions used in the main algorithm are related to tag 
lists and the CS-tree and are described in detail now.

\subsection{Connected Subgraph Tree (CS-tree)}
\label{section:cs-tree}

\begin{figure}
\centering
\includegraphics[width=\textwidth]{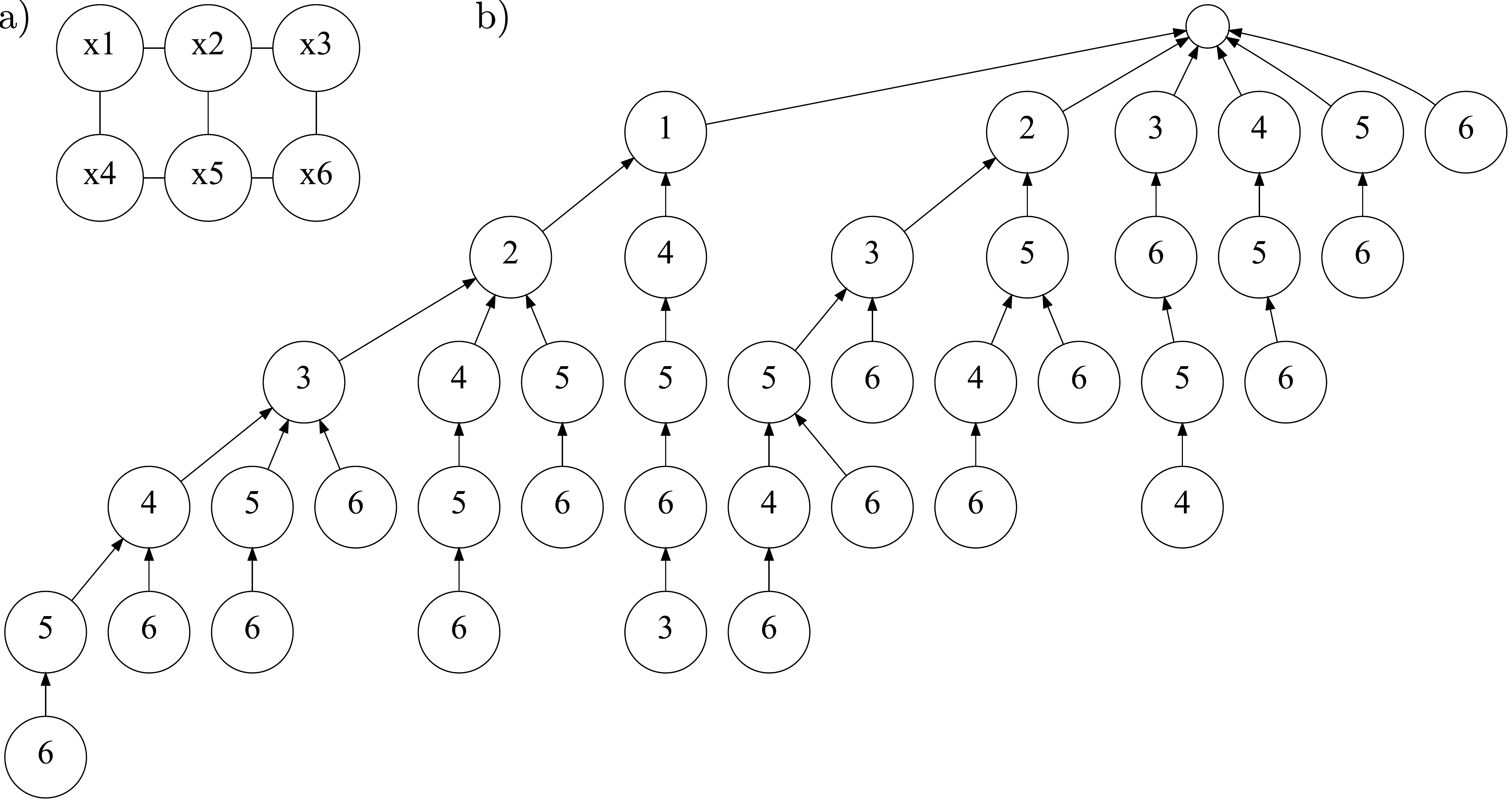}
\caption{All connected subgraphs of a graphical model (a) can be represented
uniquely in a connected subgraph tree (CS-tree) (b). Every path from a node in
the CS-tree to the root node corresponds to a connected subgraph in the
graphical model. While there are $2^6 = 64$ subsets of variables in total in this 
example, only 40 of these subsets are connected.}
\label{figure:cs-tree}
\end{figure}

The CS-tree represents subsets of connected variables uniquely. Every node in 
the CS-tree except the special root node is labeled with the integer index of 
one variable in the graphical model. The same variable index is assigned to 
several nodes in the CS-tree unless the graphical model is completely 
disconnected. The CS-tree is constructed such that every connected subset of 
variables in the graphical model corresponds to precisely one path in the 
CS-tree from a node to the root node, the node labels along the path indicating
precisely the variables in the subset, and vice versa, there exists precisely 
one connected subset of variables in the graphical model for each path in the 
CS-tree from a node to the root node. 

In order to guarantee by construction of the CS-tree that each subset of
connected variables is represented precisely once, the variable indices of each
subset are put in a special order, namely the lexicographically smallest order
in which each variable is connected to at least one of its predecessors. The
following definition of these sequences of variable indices is recursive and
therefore motivates an algorithm for the construction of the CS-tree for the
Lazy Flipper. A small grid model and its complete CS-tree are depicted in
Fig.~\ref{figure:cs-tree}.

\begin{definition}[CSR-Sequence]
\label{definition:subset-representing-sequence}
Given an undirected graph $G = (V, E)$ whose $m \in \mathbb{N}$ vertices
$V = \{1,\ldots,m\}$ are integer indices, every sequence that consists of only
one index is called \emph{connected subset representing (CSR)}. Given
$n \in \mathbb{N}$ and a CSR-sequence $(v_1, \ldots, v_n)$, a sequence
$(v_1, \ldots, v_n, v_{n+1})$ of $n+1$ indices is called a \emph{CSR-sequence}
precisely if the following conditions hold:

(i) $v_{n+1}$ is not among its predecessors,
i.e.~$\forall j \in \{1, \ldots, n\}: v_j \not= v_{n+1}$.

(ii) $v_{n+1}$ is connected to at least one of its predecessors,
i.e.~$\exists j \in \{1, \ldots, n\}: \{v_j, v_{n+1}\} \in E$.

(iii) $v_{n+1} > v_1$.

(iv) If $n \geq 2$ and $v_{n+1}$ could have been added at an earlier position
$j \in \{2,\ldots,n\}$ to the sequence, fulfilling (i)--(iii), all subsequent
vertices $v_j, \ldots, v_n$ are smaller than $v_{n+1}$, i.e.
\begin{equation}
\forall j \in \{2,\ldots,n\}
\left(
  \{v_{j-1}, v_{n+1}\} \in E
  \Rightarrow
  \left(
    \forall k \in \{j,\ldots,n\}: v_k < v_{n+1}
  \right)
\right) \enspace .
\end{equation}
\end{definition}

Based on this definition, three functions are sufficient to recursively build
the CS-tree $T$ of a graphical model $G$, starting from the root node. The
function $q$ = \emph{growSubset}($T,G,p$) appends to a node $p$ in the CS-tree
the smallest variable index that is not yet among the children of $p$ and
fulfills (i)--(iv) for the CSR-sequence of variable indices on the path from
$p$ to the root node. It returns the appended node or the empty set if no suitable
variable index exists. The function $q$ = \emph{firstSubsetOfSize}($T,G,n$)
traverses the CS-tree on the current deepest level $n-1$, calling the function
\emph{growSubset} for each leaf until a node can be appended and thus, the
first subset of size $n$ has been found. Finally, the function
$q$ = \emph{nextSubsetOfSameSize}($T,G,p$) starts from a node $p$, finds its
parent and traverses from there in level order, calling \emph{growSubset} for
each node to find the length-lexicographic successor of the CSR-sequence
associated with the node $p$, i.e.~the representative of the next subset of the
same size. These functions are used by the Lazy Flipper
(Alg.~\ref{algorithm:lazy-flipper}) to \emph{construct} the CS-tree.

In contrast, the \emph{traversal} of already constructed parts of the CS-tree
(when revisiting subsets of variables after successful flips) is performed by
functions associated with tag lists which are defined the following section.

\subsection{Tag Lists}
\label{section:tag-lists}

Tag lists are used to tag variables that are affected by flips. A variable is
affected by a flip either because it has been flipped itself or because it is
connected (via a potential) to a flipped variable. The tag list data structure
comprises a Boolean vector in which each entry corresponds to a variable,
indicating whether or not this variable is affected by recent flips. As
the total number of variables can be large ($10^6$ is not exceptional) and
possibly only a few variables are affected by flips, a list of all affected
variables is maintained in addition to the vector. This list allows the
algorithm to untag all tagged variables without re-initializing the entire
Boolean vector. The two fundamental operations on a tag list $L$ are
\emph{tag}$(L,x)$ which tags the variable with the index $x$, and
\emph{untagAll}$(L)$.

For the Lazy Flipper, three special functions are used in addition: Given a tag
list $L$, a (possibly incomplete) CS-tree $T$, the graphical model $G$, and a
node $s \in T$, $\textit{tagConnectedVariables}(L,T,G,s)$ tags all variables on
the path from $s$ to the root node in $T$, as well as all nodes that are
connected (via a potential in $G$) to at least one of these nodes. The function
$s = \textit{firstTaggedSubset}(L,T)$ traverses the first level of $T$ and
returns the first node $s$ whose variable is tagged (or the empty set if all
variables are untagged). Finally, the function
$t = \textit{nextTaggedSubset}(L,T,s)$ traverses $T$ in level order, starting
with the successor of $s$, and returns the first node $t$ for which the path to
the root contains at least one tagged variable. These functions, together with
those of the CS-tree, are sufficient for the Lazy Flipper,
Alg.~\ref{algorithm:lazy-flipper}.

\section{The Lazy Flipper Algorithm}
\label{section:algorithm}

In the main loop of the Lazy Flipper (lines 2--26 in Alg.~\ref{algorithm:lazy-flipper}),
the size $n$ of subsets is incremented until the limit
$\nmax$ is reached (line 24). Inside this main loop, the algorithm falls into two
parts, the \emph{exploration part} (lines 3--11) and the \emph{revisiting part}
(lines 12--23). In the exploration part, flips of previously unseen subsets of
$n$ variables are assessed. The current best configuration $c$ is updated in a
greedy fashion, i.e.~whenever a flip yields a lower energy. At the same time, the
CS-tree is grown, using the functions defined in Section \ref{section:cs-tree}.
In the revisiting part, all subsets of sizes $1$ through $n$ that are
affected by recent flips are assessed iteratively until no flip of any of these 
subsets reduces the energy (line 14). The indices of affected variables are stored
in the tag lists $L_1$ and $L_2$ (cf.~Section \ref{section:tag-lists}).
In practice, the Lazy Flipper can be stopped at any point, e.g.~when a time limit
is exceeded, and the current best configuration $c$ taken as the output. It
eventually reaches configurations for which it is guaranteed that no flip of
$n$ or less variables can yield a lower energy because all such flips that could 
potentially lower the energy have been assessed (line 14). Such configurations are
therefore guaranteed to be optimal within a Hamming radius of $n$:

\begin{definition}[Hamming-$n$ bound]
Given a function $E: \{0,1\}^m \rightarrow \mathbb{R}$, a configuration
$c \in \{0,1\}^m$, and $n \in \mathbb{N}$, $E(c)$ is called a \emph{Hamming-}$n$
\emph{upper bound} on the minimum of $E$ precisely if
$\forall c' \in \{0,1\}^m ( |c' - c|_1 \leq n \Rightarrow E(c) \leq E(c') )$.
\end{definition}

\begin{algorithm}[h!]
\caption{Lazy Flipper}
\label{algorithm:lazy-flipper}
\KwIn{$G$: graphical model with $m \in \mathbb{N}$ binary variables,
$c \in \{0,1\}^m$: initial configuration,
$\nmax \in \mathbb{N}$: maximum size of subgraphs to be searched}
\KwOut{$c \in \{0,1\}^m$ (modified): configuration corresponding to the
smallest upper bound found ($c$ is optimal within a Hamming radius of $\nmax$).}
$n \leftarrow 1$;
CS-Tree $T \leftarrow \{$root$\}$;
TagList $L_1 \leftarrow \emptyset$, $L_2 \leftarrow \emptyset$\;
\Forever{
  $s \leftarrow$ firstSubsetOfSize$(T, G, n)$\;
  \lIf{$s = \emptyset$}{%
    \Break \;
  }

  \While{$s \not= \emptyset$}{
    \If{energyAfterFlip$(G, c, s) <$ energy($G, c$)}{
      $c \leftarrow$ flip$(c, s)$\;
      tagConnectedVariables$(L_1, T, G, s)$\;
    }
    $s \leftarrow$ nextSubsetOfSameSize$(T, G, s)$\;
  }
  \Forever{
    $s_2 \leftarrow$ firstTaggedSubset$(L_1, T)$\;
    \lIf{$s_2 = \emptyset$}{%
      \Break\;
    }
    \While{$s_2 \not= \emptyset$}{
      \If{energyAfterFlip($G, c, s_2$) $<$ energy($G, c$)}{
        $c \leftarrow$ flip$(c, s_2)$\;
        tagConnectedVariables$(L_2, T, G, s_2)$\;
      }
      $s_2 \leftarrow$ nextTaggedSubset$(L_1, T, s_2)$\;
    }
    untagAll$(L_1)$; 
    swap$(L_1, L_2)$\;
  }
  \lIf{$n = \nmax$}{%
    \Break\;
  }
  $n \leftarrow n + 1$\;
}
\end{algorithm}

\section{Experiments}
\label{section:experiments}

For a comparative assessment of the Lazy Flipper, four optimization problems of
different complexity are considered, two simulated problems and two problems 
based on real-world data. For the sake of reproducibility, the simulations are 
described in detail and the models constructed from real data are available 
from the authors as supplementary material.

The first problem is a ferromagnetic Ising model that is widely used in 
computer vision for foreground vs.~background segmentation 
\cite{boycov-2001}. 
Energy functions of this model consist of first and second order potentials 
that are submodular. The global minimum can therefore be found via a graph cut.
We simulate random instances of this model in order to measure how the runtime 
of lazy flipping depends on the size of the model and the coupling strength, 
and to compare Lazy Flipper approximations to the global optimum
(Section~\ref{section:ising-model}).

The second problem is a problem of finding optimal subgraphs on a grid. 
Energy functions of this model consist of first and fourth order potentials, 
of which the latter are not permutation submodular. We simulate difficult 
instances of this problem that cannot be solved to optimality, even when 
allowing several days of runtime. In this challenging setting, Lazy Flipper 
approximations and their convergence are compared to those of BP, TRBP and DD 
as well as to the lower bounds on local polytope relaxations obtained by DD
(Section~\ref{section:optimal-subgraph-model}).

The third problem is a graphical model for removing excessive boundaries from 
image over-segmentations that is related to the model proposed in 
\cite{zhang-2010}.
Energy functions of this model consist of first, third and fourth order 
potentials. In contrast to the grid graphs of the Ising model and the optimal 
subgraph model, the corresponding factor graphs are irregular but still planar. 
The higher-order potentials are not permutation submodular but the global 
optimum can be found by means of MILP in approximately 10 seconds per model
using one of the fastest commercial solvers (IBM ILOG CPLEX, version~12.1). 
Since CPLEX is closed-source software, the algorithm is not known in detail 
and we use it as a black box. The general method used by CPLEX for MILP is a 
branch-and-bound algorithm
\cite{dakin-1965,land-1960}.
100 instances of this model obtained from the 100 natural test images of the 
Berkeley Segmentation Database (BSD)
\cite{martin-2001}
are used to compare the Lazy Flipper to algorithms based on message passing and
linear programming in a real-world setting where the global optimum is 
accessible
(Section~\ref{section:2d-segmentation-model}).

The fourth problem is identical to the third, except that instances are 
obtained from 3-dimensional volume images of neural tissue acquired by means of
Serial Block Face Scanning Electron Microscopy (SBFSEM)
\cite{denk-2004}.
Unlike in the 2-dimensional case, the factor graphs are no longer planar. 
Whether exact optimization by means of MILP is practical depends on the size of 
the model. In practice, SBFSEM datasets consist of more than 2000$^3$ voxels. To 
be able to compare approximations to the \emph{global} optimum, we consider 16 
models obtained from 16 SBFSEM volume sub-images of only 150$^3$ voxels for 
which the global optimum can be found by means of MILP within a few minutes 
(Section~\ref{section:3d-segmentation-model}).

\subsection{Ferromagnetic Ising model}
\label{section:ising-model}

The ferromagnetic Ising model consists of $m \in \mathbb{N}$ binary variables
$x_1,\ldots,x_m \in \{0,1\}$ that are associated with points on a 2-dimensional 
square grid and connected via second order potentials 
$E_{jk}(x_j, x_k) = 1-\delta_{x_j,x_k}$
($\delta$: Kronecker delta) to their nearest neighbors. First order
potentials $E_j(x_j)$ relate the variables to observed evidence in underlying
data. The total energy of this model is the following sum in which $\alpha \in
\mathbb{R}_0^+$ is a weight on the second order potentials, and $j \sim k$ 
indicates that the variables $x_j$ and $x_k$ are adjacent on the grid:
\begin{equation}
\forall x \in \{0,1\}^m:
\quad
E(x) = \sum_{j=1}^{m}{ E_j(x_j) }
+ \alpha \sum_{j=1}^{m}{
    \sum_{ \substack{k=j+1 \\ k \sim j} }^{m}{
      E_{jk}(x_j, x_k)
    }
  }
\enspace .
\end{equation}
For each $\alpha \in \{0.1, 0.3, 0.5, 0.7, 0.9\}$, an ensemble of ten simulated
Ising models of $50 \cdot 50 = 2500$ variables is considered. The first order 
potentials $E_j$ are initialized randomly by drawing $E_j(0)$ uniformly from 
the interval $[0,1]$ and setting $E_j(1) := 1-E_j(0)$. The exact global minimum 
of the total energy is found via a graph cut.

\begin{figure}
\includegraphics[width=0.49\textwidth]{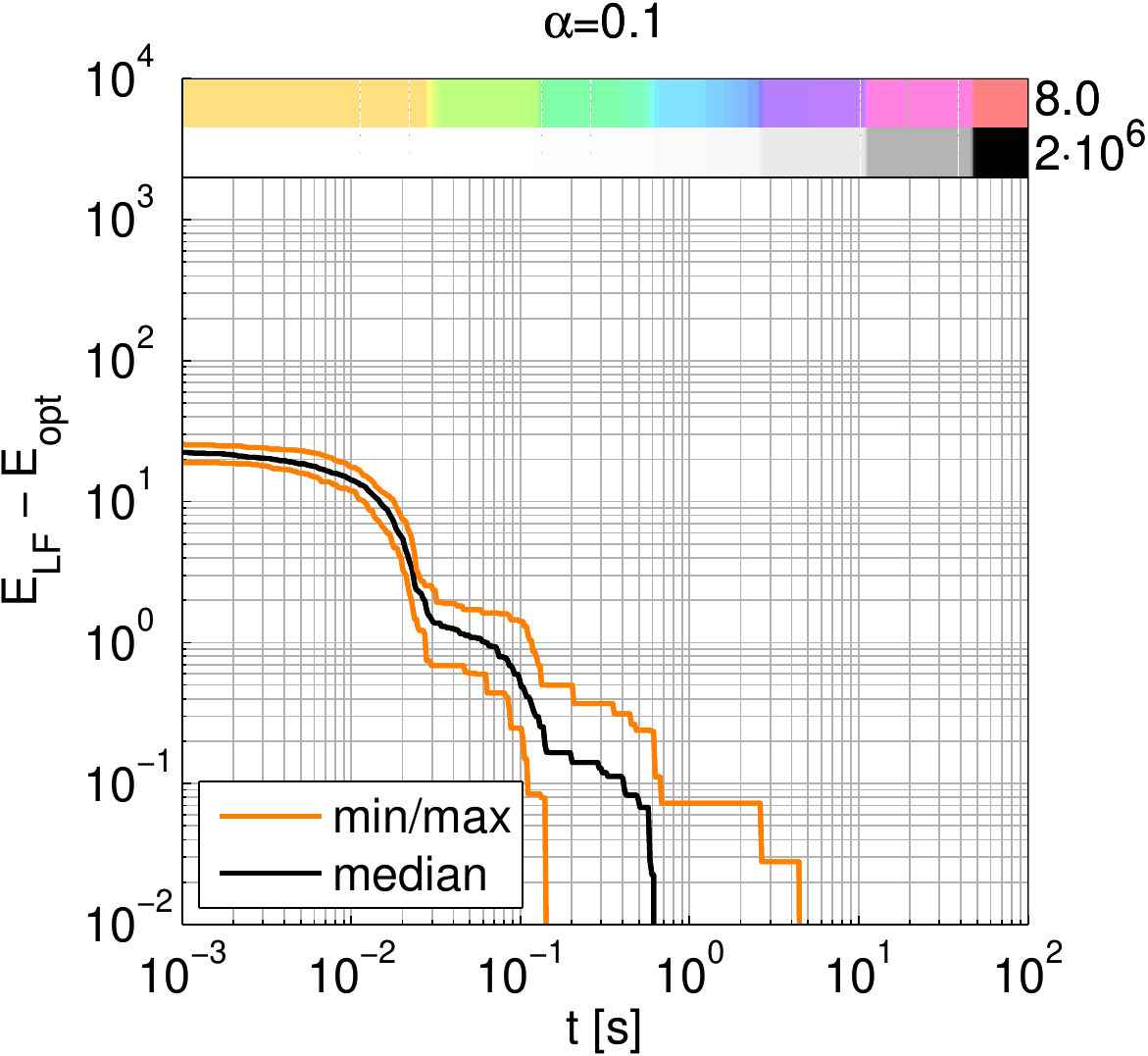}
\includegraphics[width=0.49\textwidth]{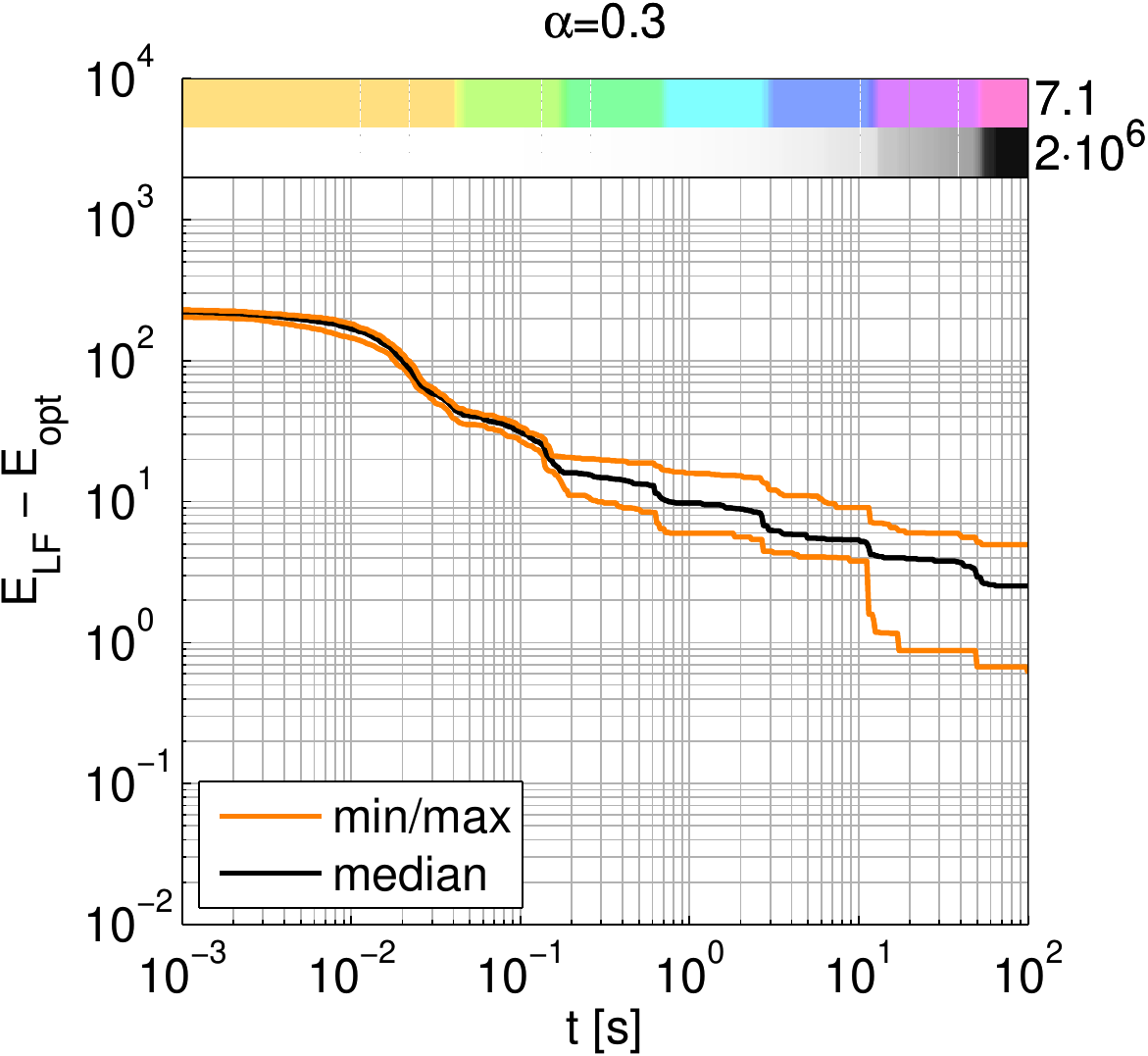}
\vspace{2ex}\\
\includegraphics[width=0.49\textwidth]{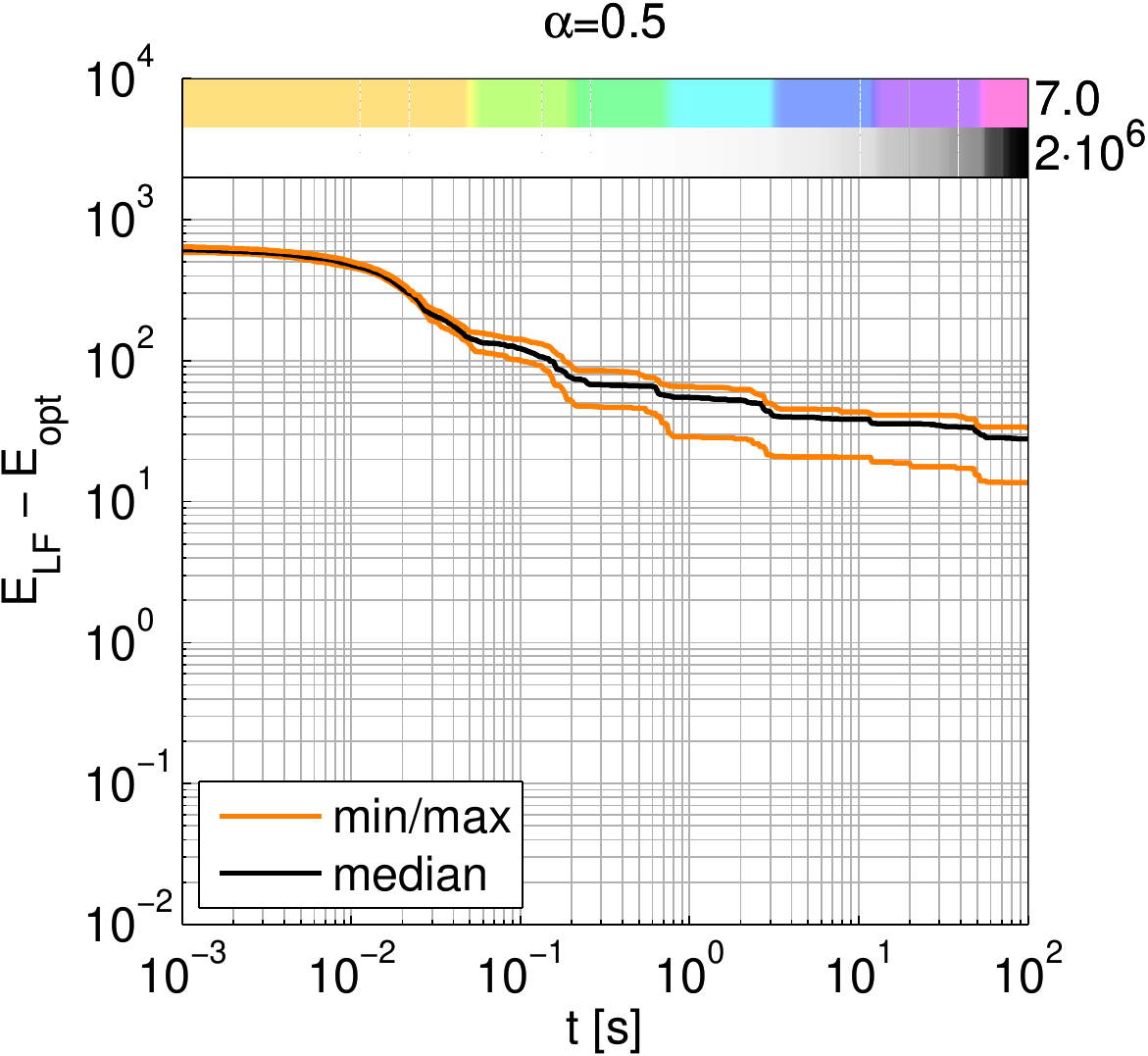}
\includegraphics[width=0.49\textwidth]{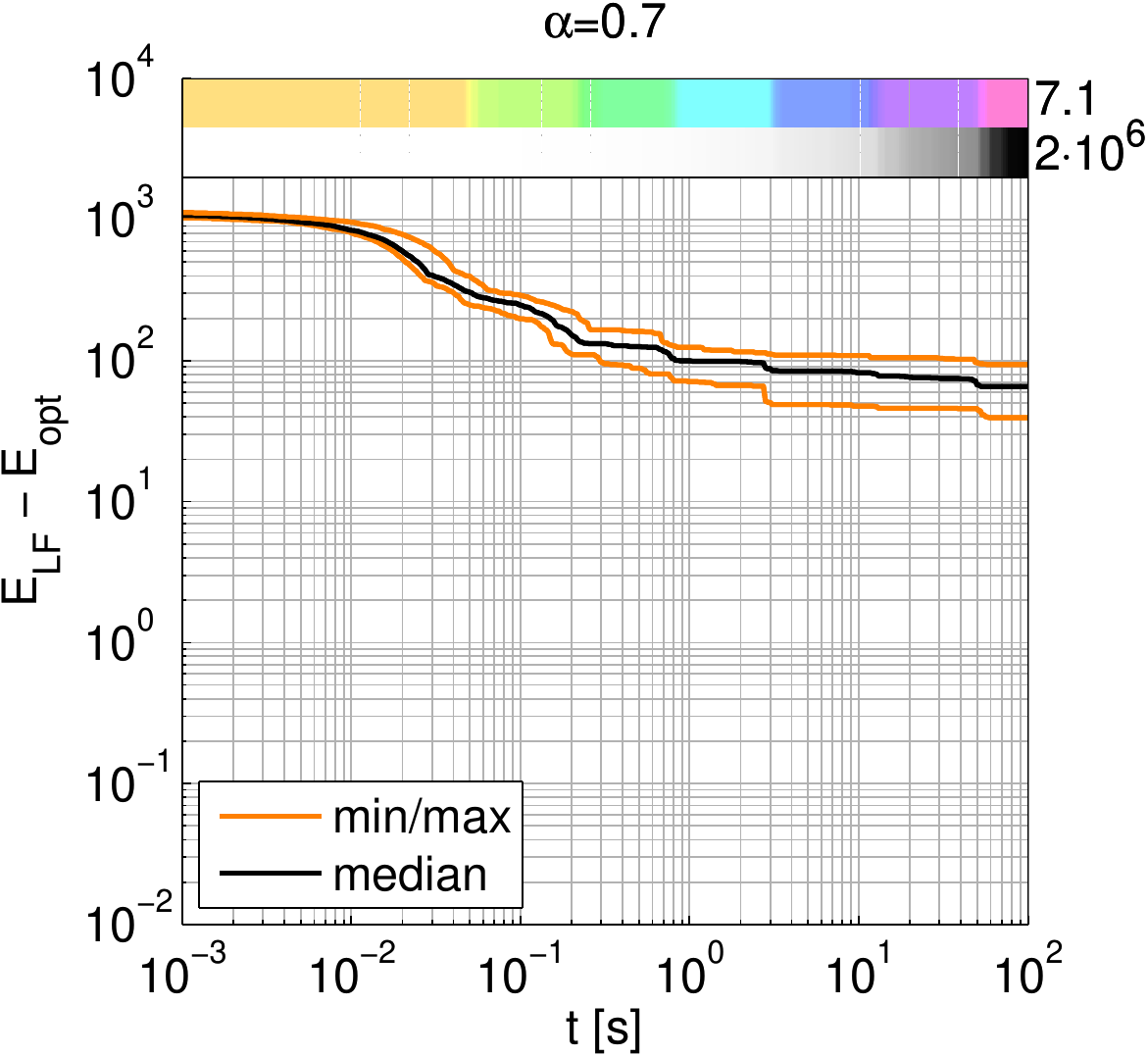}
\vspace{2ex}\\
\includegraphics[width=0.49\textwidth]{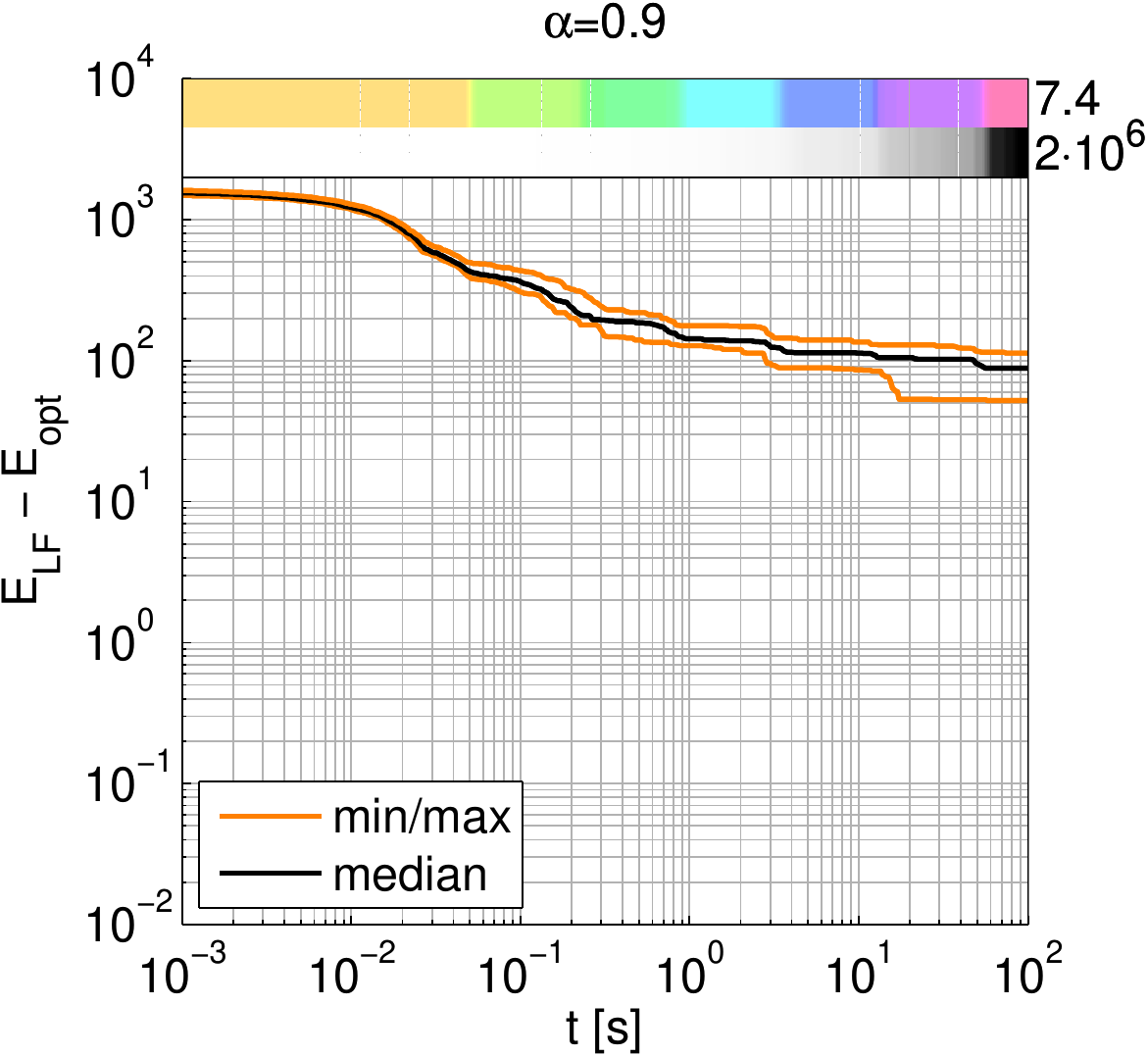}
\includegraphics[width=0.45\textwidth]{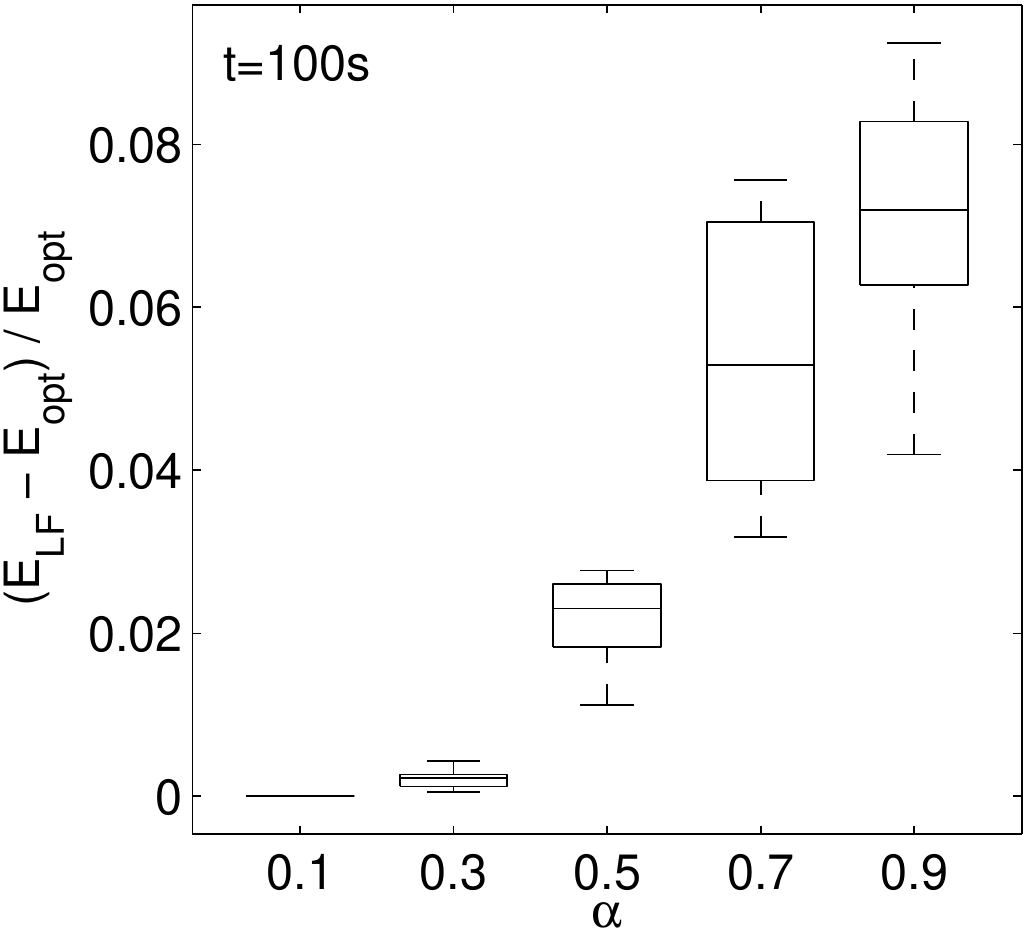}
\caption{Upper bounds on the minimum energy of a graphical model can be found by 
flipping subsets of variables. The deviation of these upper bounds from the 
minimum energy is shown above for ensembles of ten random Ising models 
(Section~\ref{section:ising-model}). 
Compared to optimization by ICM where only one variable is flipped at a time, 
the Lazy Flipper finds significantly tighter bounds by flipping also larger 
subsets. The deviations increase with the coupling strength $\alpha$. Color 
scales and gray scales indicate the size and the total number of searched 
subsets.}
\label{figure:ising-results}
\end{figure}

For each model, the Lazy Flipper is initialized with a configuration that 
minimizes the sum of the first order potentials. Upper bounds on the minimum 
energy found by means of lazy flipping converge towards the global optimum as 
depicted in 
Fig.~\ref{figure:ising-results}. 
Color scales and gray scales in this figure respectively indicate the maximum 
size and the total number of distinct subsets that have been searched, averaged 
over all models in the ensemble. It can be seen from this figure that upper 
bounds on the minimum energy are tightened significantly by searching larger 
subsets of variables, independent of the coupling strength $\alpha$. It takes 
the Lazy Flipper less than 100~seconds (on a single CPU of an Intel Quad Xeon 
E7220 at 2.93GHz) to exhaustively search all connected subsets of 6~variables. 
The amount of RAM required for the CS-tree (in bytes) is 24~times as high as 
the number of subsets (approximately 50~MB in this case) because each subset is
stored in the CS-tree as a node consisting of three 64-bit integers: a variable
index, the index of the parent node and the index of the level order successor
(Section~\ref{section:cs-tree})\footnote{The size of the CS-tree becomes 
limiting for very large problems. However, for regular graphs, implicit 
representations can be envisaged that overcome this limitation.}.

\begin{figure}
\centering
\begin{tabular}{cccccc}
$\nmax$ & $\alpha=0.1$ & $\alpha=0.3$ & $\alpha=0.5$ & $\alpha=0.7$ & $\alpha=0.9$\\
$1$ &
\includegraphics[width=2cm]{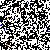} &
\includegraphics[width=2cm]{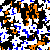} &
\includegraphics[width=2cm]{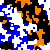} &
\includegraphics[width=2cm]{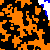} &
\includegraphics[width=2cm]{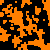} \\
$6$ &
\includegraphics[width=2cm]{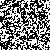} &
\includegraphics[width=2cm]{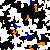} &
\includegraphics[width=2cm]{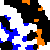} &
\includegraphics[width=2cm]{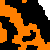} &
\includegraphics[width=2cm]{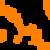} \\
$\infty$ &
\includegraphics[width=2cm]{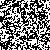} &
\includegraphics[width=2cm]{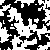} &
\includegraphics[width=2cm]{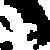} &
\includegraphics[width=2cm]{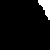} &
\includegraphics[width=2cm]{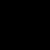} \\
\end{tabular}
\caption{The configurations found by the Lazy Flipper converge to a global
optimum as the search depth $\nmax$ increases. For Ising models with different 
coupling strengths $\alpha$ (columns), deviations from the global optimum 
($\nmax = \infty$) are depicted in blue (false 0) and orange (false 1), for 
$\nmax \in \{1,6\}$. As the Lazy Flipper is greedy, these approximate solutions 
highly depend on the initialization and on the order in which subsets are visited.}
\label{figure:ising-states}
\end{figure}

\begin{figure}
\centering
\includegraphics[height=6cm]{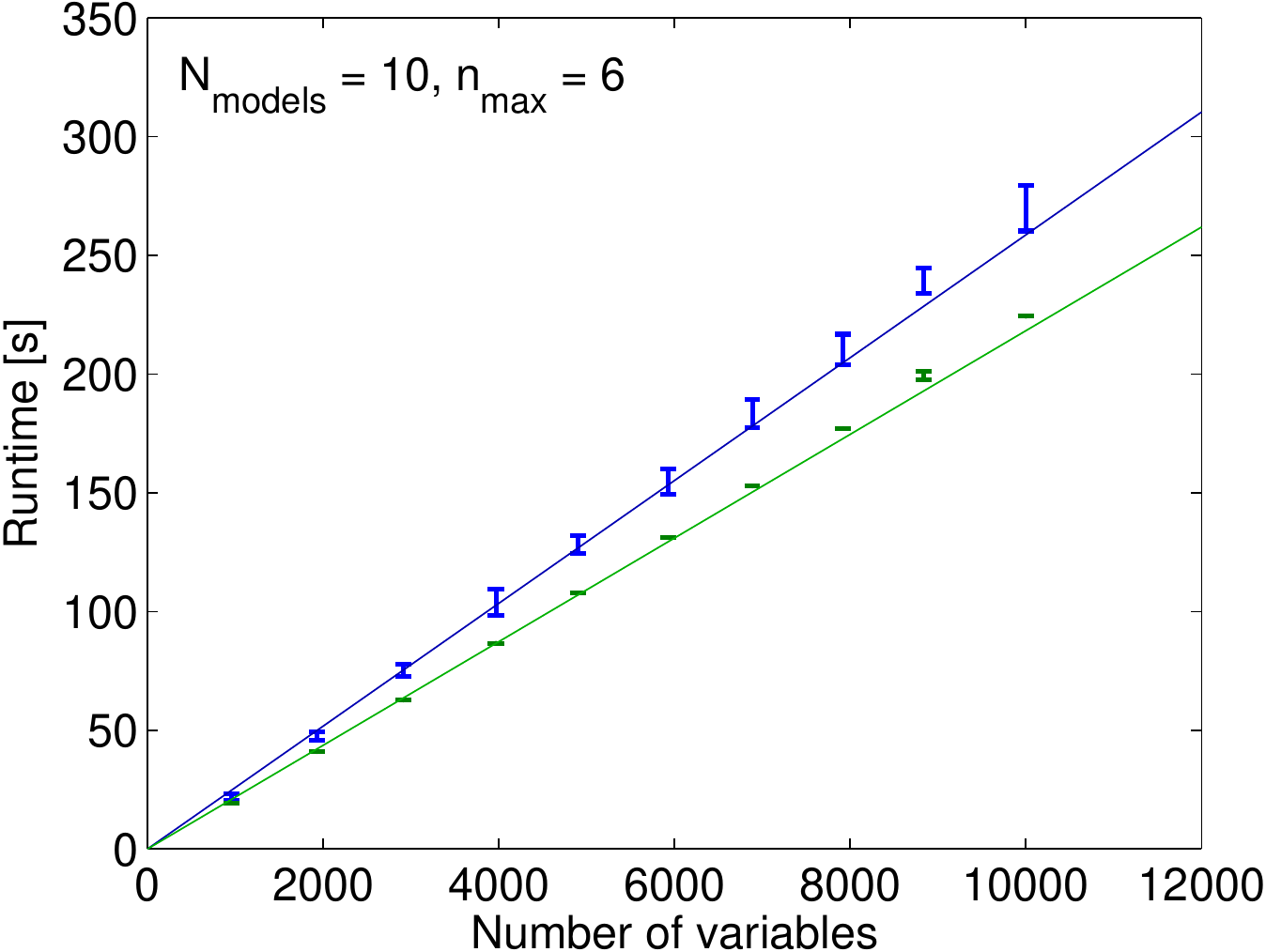}
\caption{For a fixed maximum subgraph size ($\nmax = 6$), the runtime of lazy 
flipping scales only slightly more than linearly with the number of variables 
in the Ising model. It is measured for coupling strengths $\alpha=0.25$ 
(upper curve) and $\alpha=0.75$ (lower curve). Error bars indicate the standard
deviation over 10 random models, and lines are fitted by least squares. Lazy 
flipping takes longer ($0.0259$ seconds per variable) for $\alpha=0.25$ than 
for $\alpha=0.75$ ($0.0218$~s/var) because more flips are successful and thus 
initiate revisiting.}
\label{figure:runtime-scaling}
\end{figure}

For $\nmax \in \{1,6\}$, configurations corresponding to the upper bounds on the
minimum energy are depicted in 
Fig.~\ref{figure:ising-states}. 
It can be seen from this figure that all connected subsets of falsely set 
variables are larger than $\nmax$.
For a fixed maximum subgraph size $\nmax$, the runtime of lazy flipping scales 
approximately linearly with the number of variables in the Ising model 
(cf.~Fig.\ref{figure:runtime-scaling}).

\subsection{Optimal Subgraph Model}
\label{section:optimal-subgraph-model}

The optimal subgraph model consists of $m \in \mathbb{N}$ binary variables 
$x_1,\ldots,x_m \in \{0,1\}$ that are associated with the edges of a 
2-dimensional grid graph. A subgraph is defined by those edges whose associated 
variables attain the value 1. Energy functions of this model consist of first 
order potentials, one for each edge, and fourth order potentials, one for each 
node $v \in V$ in which four edges $(j,k,l,m) = \mathcal{N}(v)$ meet: 
\begin{equation}
\forall x \in \{0,1\}^m:
\quad
E(x) = 
\sum_{j=1}^{m}{ E_j(x_j) }
\ + 
\hspace{-6mm}
\sum_{(j,k,l,m) \in \mathcal{N}(V)}{
	\hspace{-6mm}
  E_{j k l m}(x_j, x_k, x_l, x_m)
}
\enspace .
\end{equation}
All fourth order potentials are equal, penalizing dead ends and branches of 
paths in the selected subgraph:
\begin{equation}
E_{j k l m}(x_j, x_k, x_l, x_m) =
\begin{cases}
0.0    & \textnormal{if}\ s = 0\\
100.0  & \textnormal{if}\ s = 1\\
0.6    & \textnormal{if}\ s = 2\\
1.2    & \textnormal{if}\ s = 3\\
2.4    & \textnormal{if}\ s = 4
\end{cases}
\quad \textnormal{with} \quad
s = x_j + x_k + x_l + x_m\ .
\end{equation}

An ensemble of 16 such models is constructed by drawing the unary potentials at
random, exactly as for the Ising models. Each model has 19800 variables, the 
same number of first order potentials, and 9801 fourth order potentials. 
Approximate optimal subgraphs are found by Min-Sum Belief Propagation (BP) with 
parallel message passing 
\cite{pearl-1988,kschischang-2001} 
and message damping 
\cite{murphy-1999},
by Tree-reweighted Belief Propagation (TRBP)
\cite{wainwright-2008}, 
by Dual Decomposition (DD) 
\cite{komodakis-2010,kappes-2010}
and by lazy flipping (LF). 
DD affords also lower bounds on the minimum energy. Details on the parameters 
of the algorithms and the decomposition of the models are given in 
Appendix~\ref{section:parameters}.

\begin{figure}
\centering
\begin{tabular}{ll}
\includegraphics[height=5.6cm]{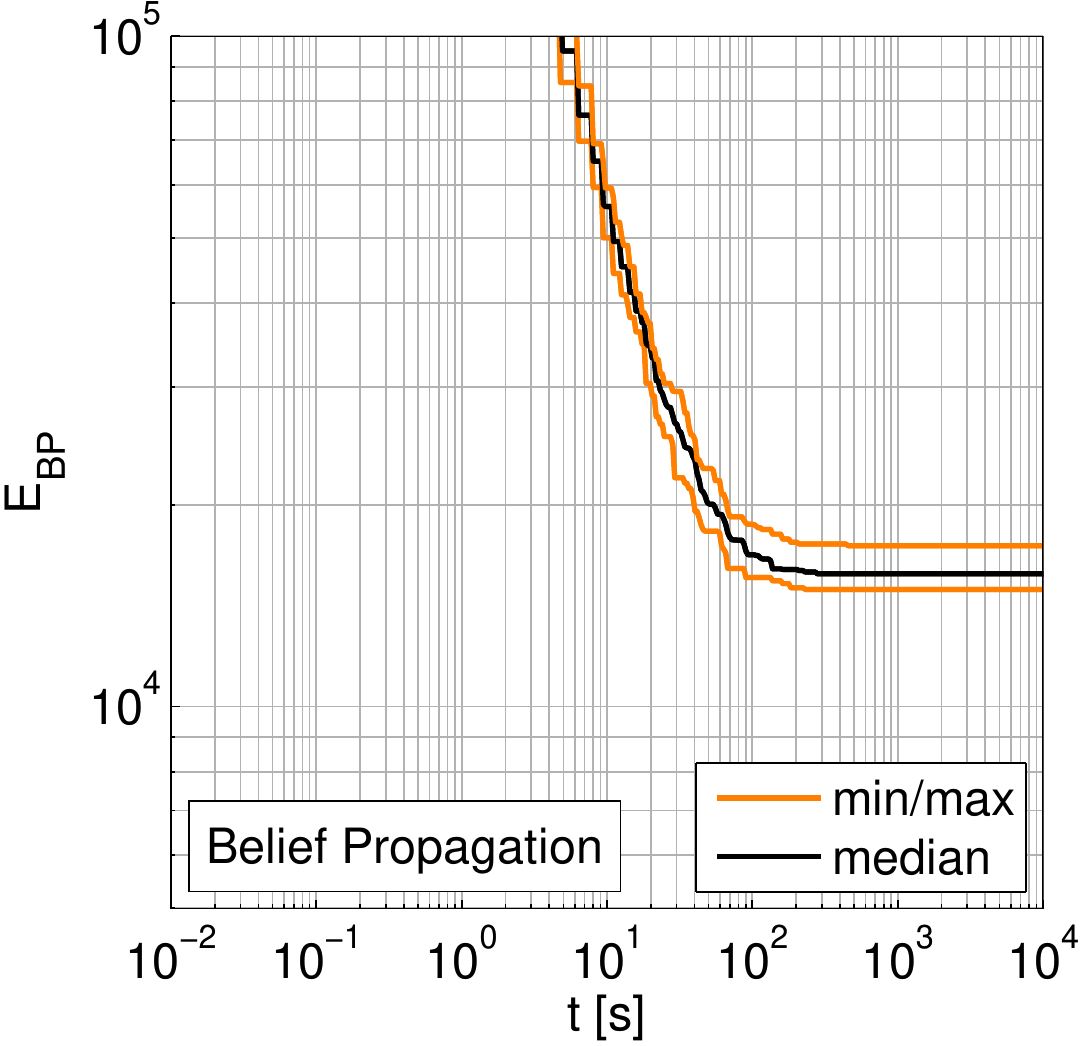} & 
\includegraphics[height=5.6cm]{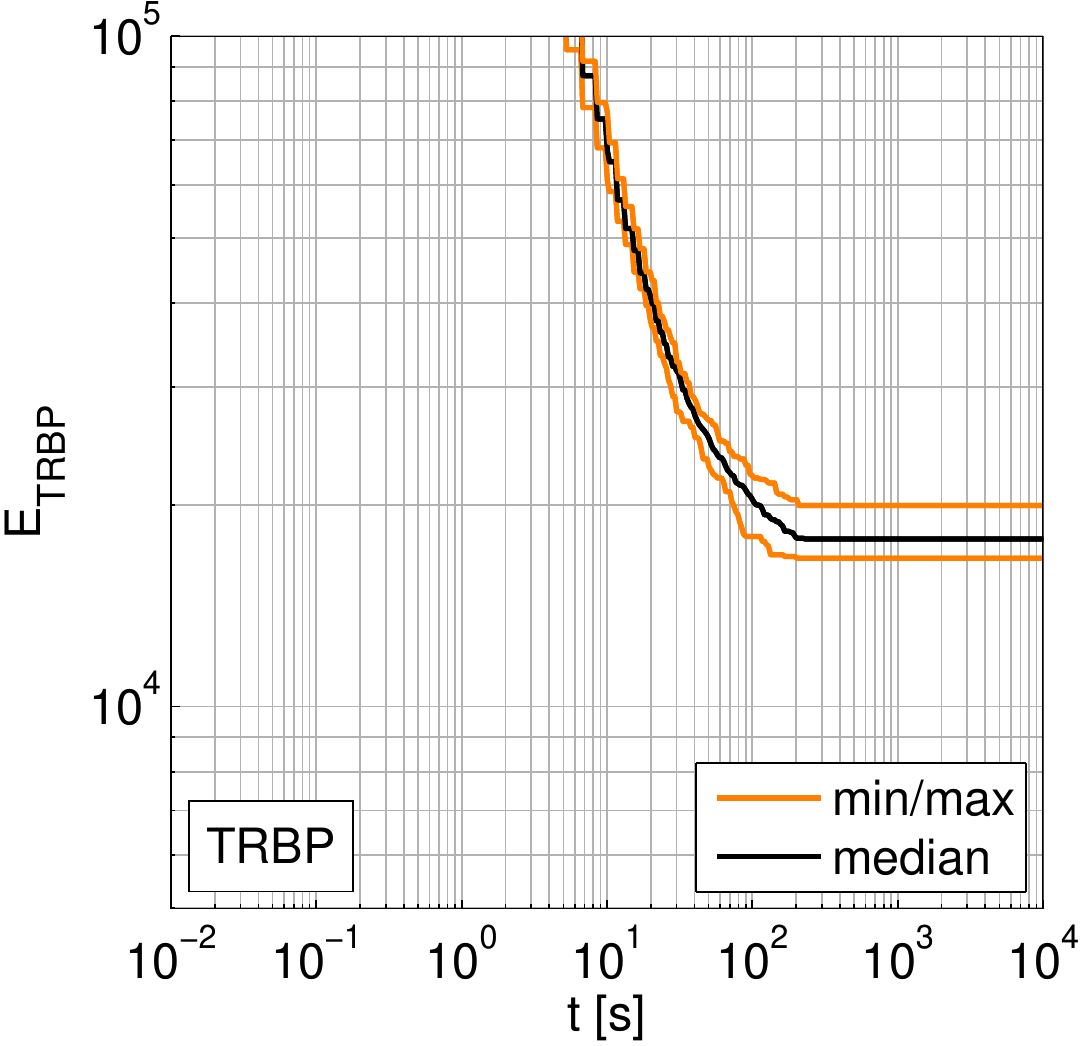} \\
\includegraphics[height=5.6cm]{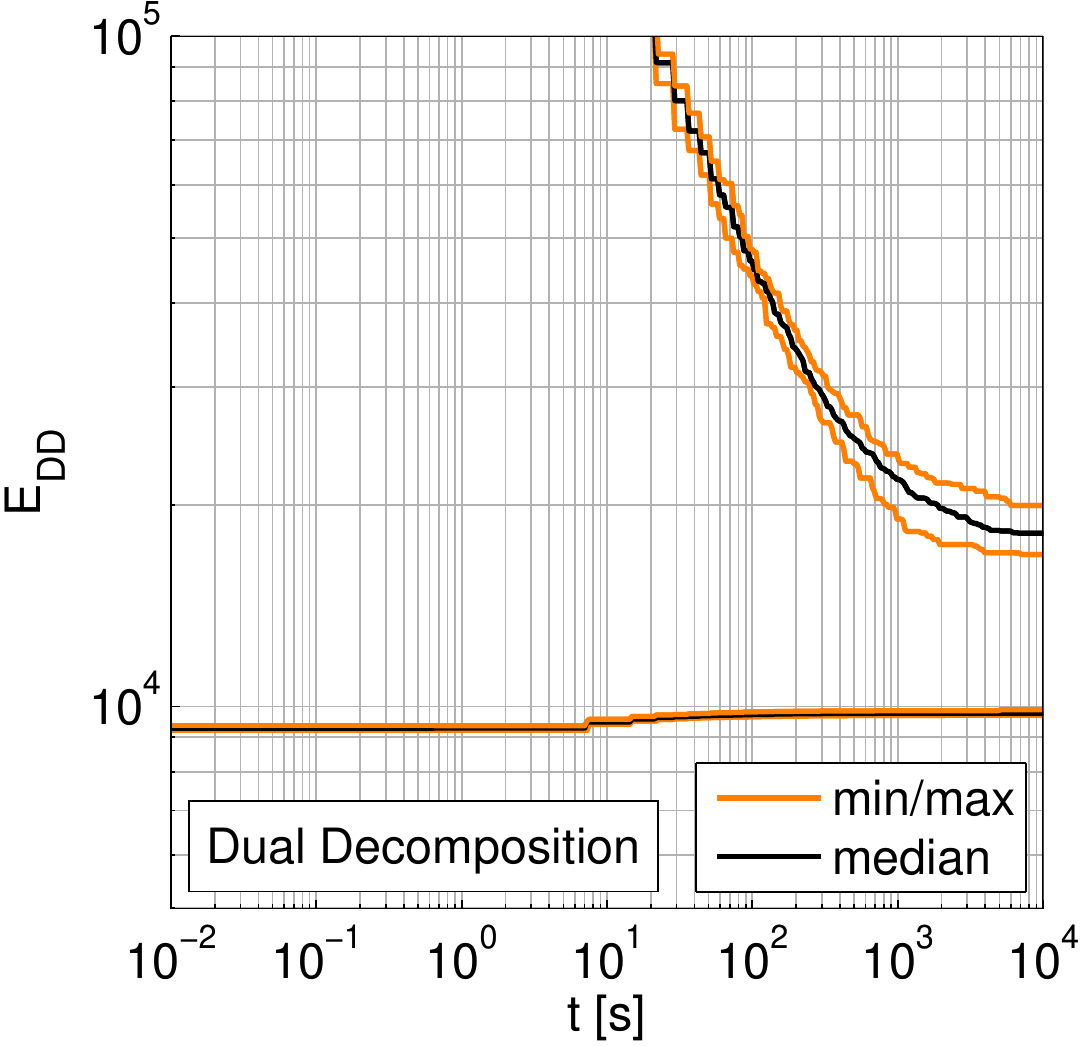} &
\includegraphics[height=5.6cm]{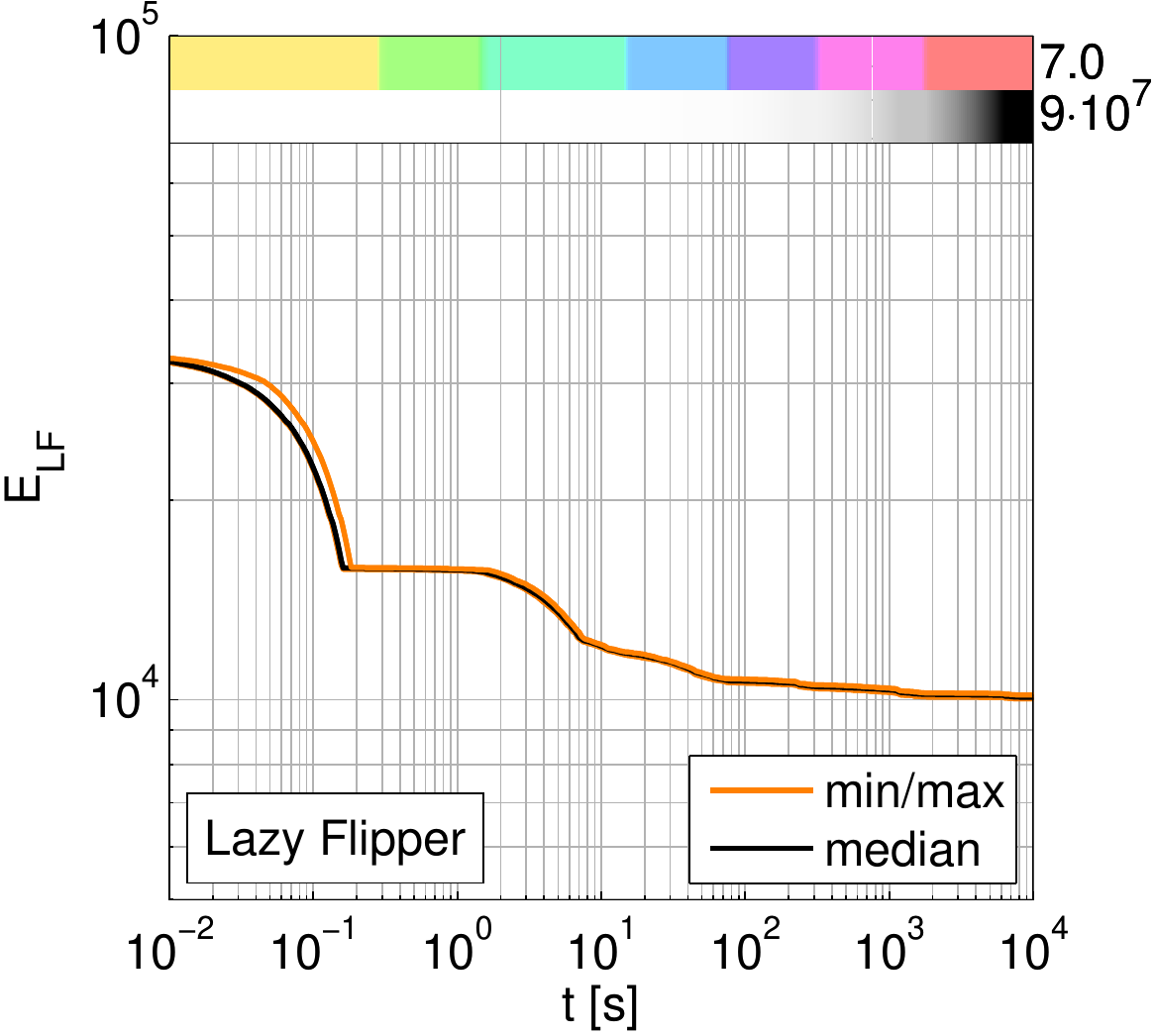}
\end{tabular}
\includegraphics[height=4.8cm]{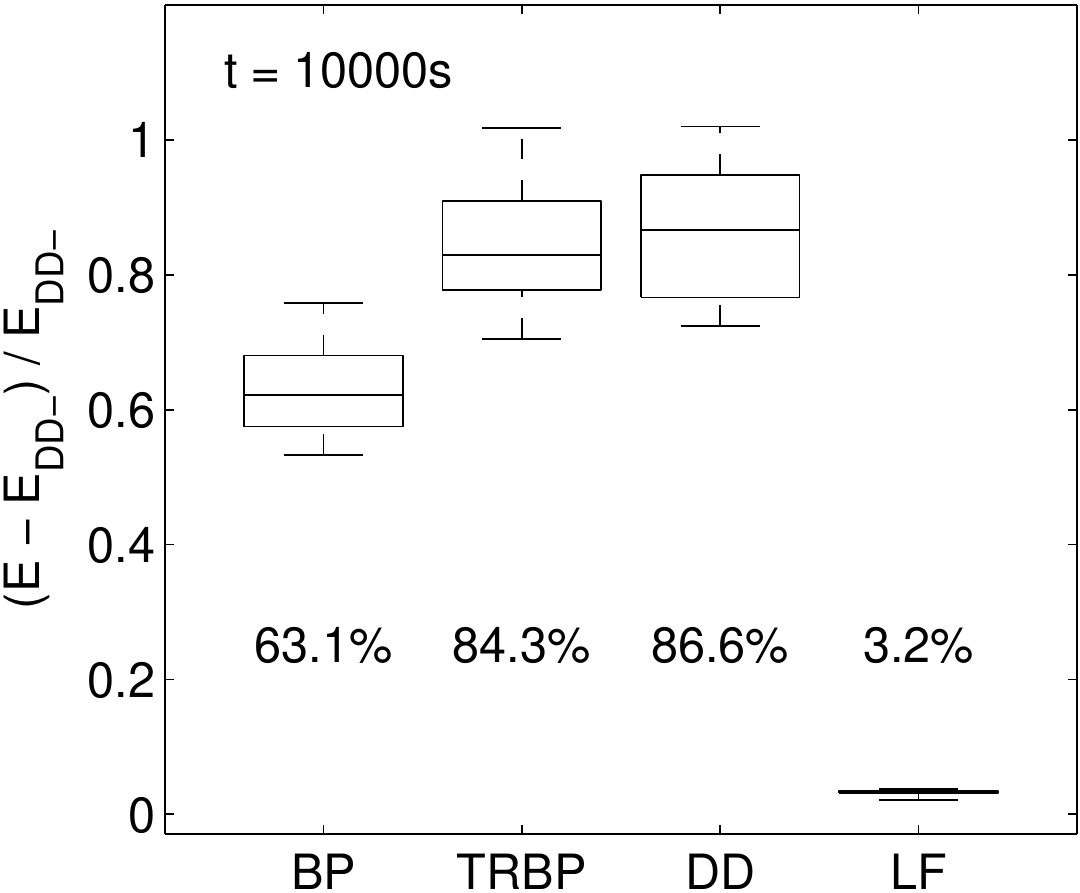}
\caption{Approximate solutions to the optimal subgraph problem 
(Section~\ref{section:optimal-subgraph-model})
are found by BP, TRBP, DD and the Lazy Flipper (LF). Depicted are the median, 
minimum and maximum (over 16 models) of the corresponding energies. DD affords 
also lower bounds on the minimum energy. The mean search depth of LF ranges 
from 1 (yellow) to 7 (red). At $t=10^4$~s, the energies of LF approximations 
come close to the lower bounds obtained by DD and thus, to the global optimum.}
\label{figure:subgraph-problem-curves}
\end{figure}

Bounds on the minimum energy converge with increasing runtime, as depicted in 
Fig.~\ref{figure:subgraph-problem-curves}. 
It can be seen from this figure that Lazy Flipper approximations converge 
fast, reaching a smaller energy after 3~seconds than the other approximations 
after 10000~seconds. Subgraphs of up to 7 variables are searched, using 
approximately 2.2~GB of RAM for the CS-tree. A gap remains between the energies 
of all approximations and the lower bound on the minimum energy obtained by DD. 
Thus, there is no guarantee that any of the problems has been solved to 
optimality. However, the gaps are upper bounds on the deviation from the global 
optimum. They are compared at $t=10000$~s in
Fig.~\ref{figure:subgraph-problem-curves}.
For any model in the ensemble, the energy of the Lazy Flipper approximation is
less than 4\% away from the global optimum, a substantial improvement over the 
other algorithms for this particular model.

\subsection{Pruning of 2D Over-Segmentations}
\label{section:2d-segmentation-model}

The graphical model for removing excessive boundaries from image 
over-segmen\-tations contains one binary variable for each boundary between 
segments, indicating whether this boundary is to be removed (0) or preserved 
(1). First order potentials relate these variables to the image content, and 
non-submodular third and fourth order potentials connect adjacent boundaries, 
supporting the closedness and smooth continuation of preserved boundaries. The 
energy function is a sum of these potentials: $\forall x \in \{0,1\}^m$
\begin{equation}
E(x) = \sum_{j=1}^{m}{ E_j(x_j) }
+ \hspace{-3mm} \sum_{(j,k,l) \in J}{ \hspace{-3mm}
    E_{jkl}(x_j, x_k, x_l)
  }
+ \hspace{-3mm} \sum_{(j,k,l,p) \in K}{ \hspace{-4mm}
    E_{jklp}(x_j, x_k, x_l, x_p)
  } \enspace .
\end{equation}

We consider an ensemble of 100 such models obtained from the 100 BSD test 
images 
\cite{martin-2001}. 
On average, a model has $8845 \pm 670$ binary variables, the same
number of unary potentials, $5715 \pm 430$ third order potentials and 
$98 \pm 18$ fourth order potentials. Each variable is connected via 
potentials to at most six other variables, a sparse structure that is favorable
for the Lazy Flipper.

\begin{figure}
\centering
\begin{tabular}{ll}
\includegraphics[height=5.6cm]{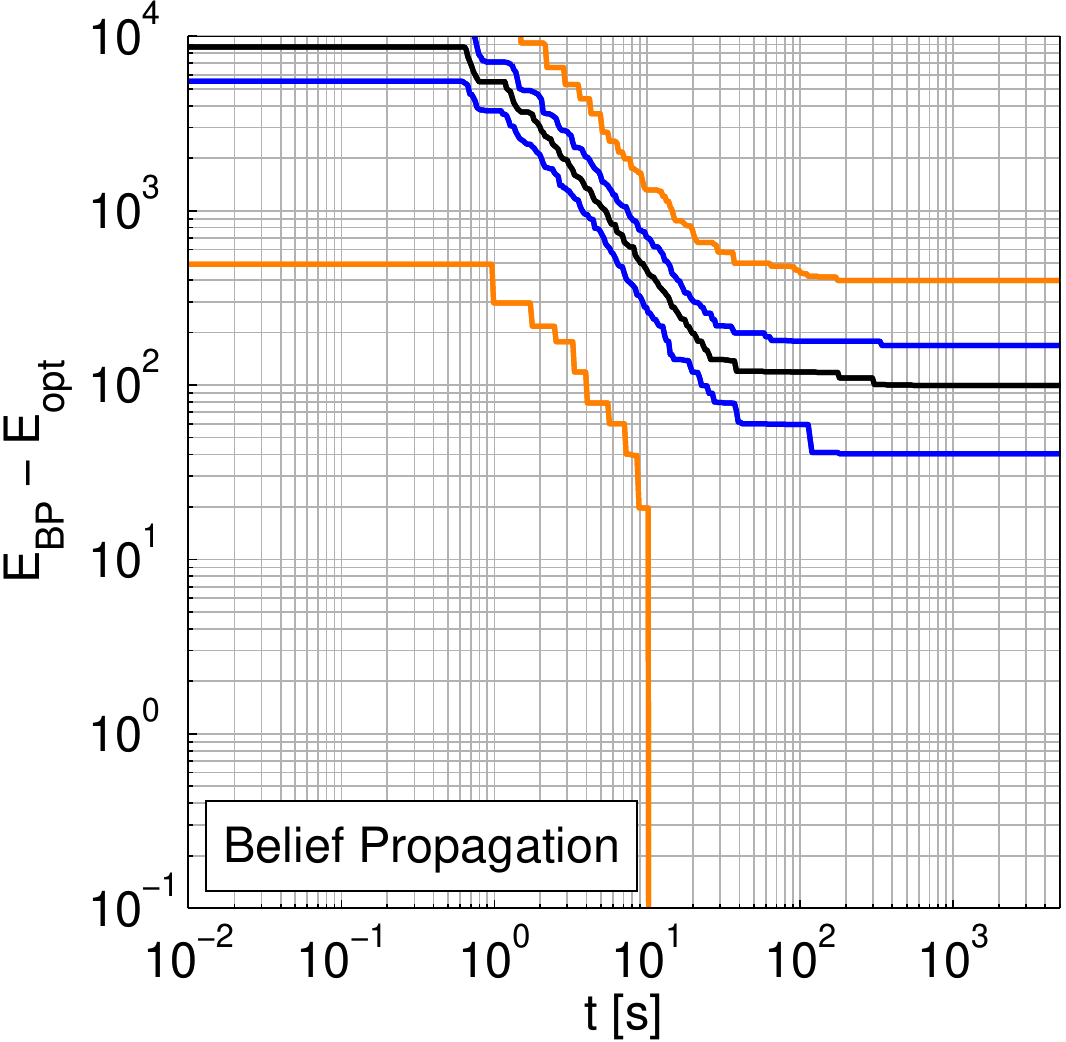} & 
\includegraphics[height=5.6cm]{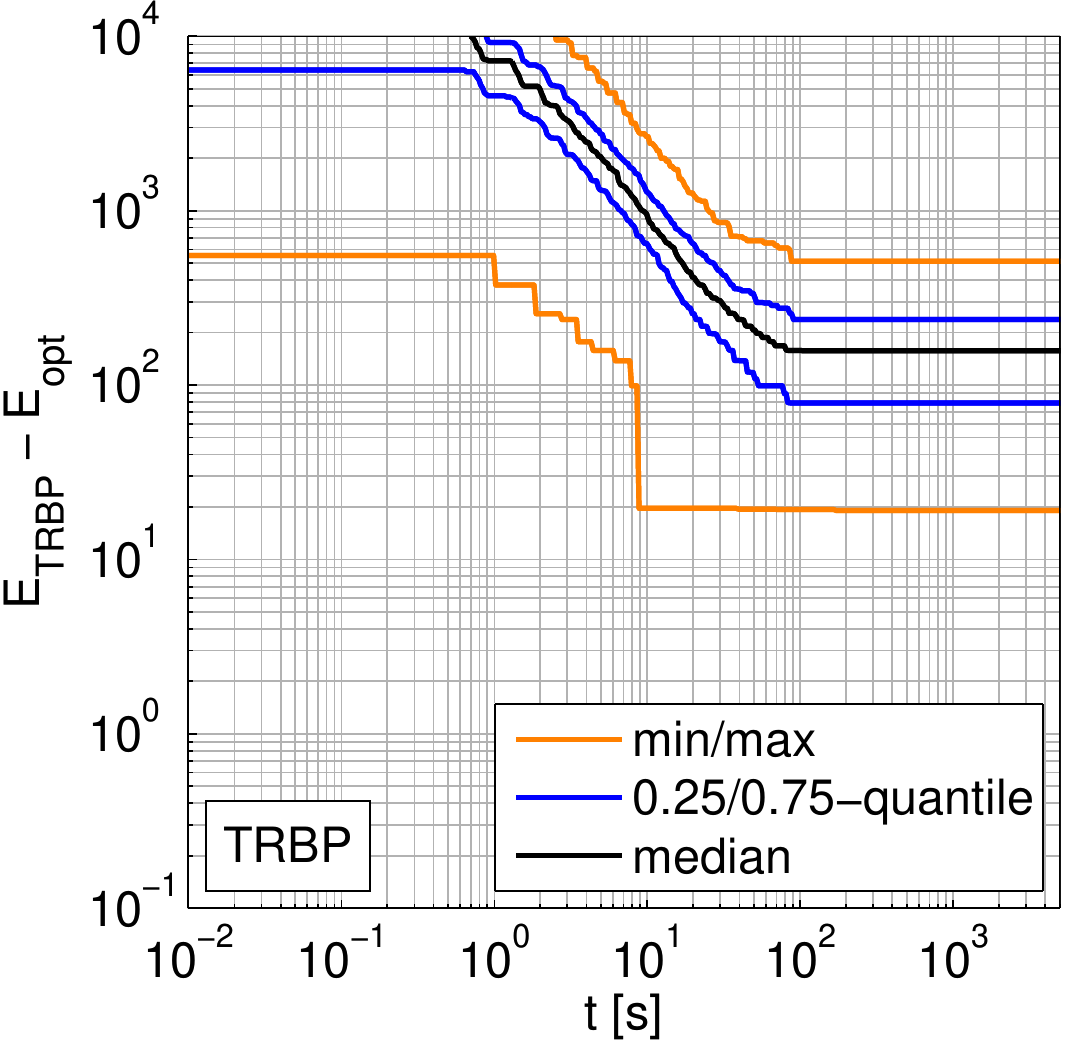} \\
\includegraphics[height=5.6cm]{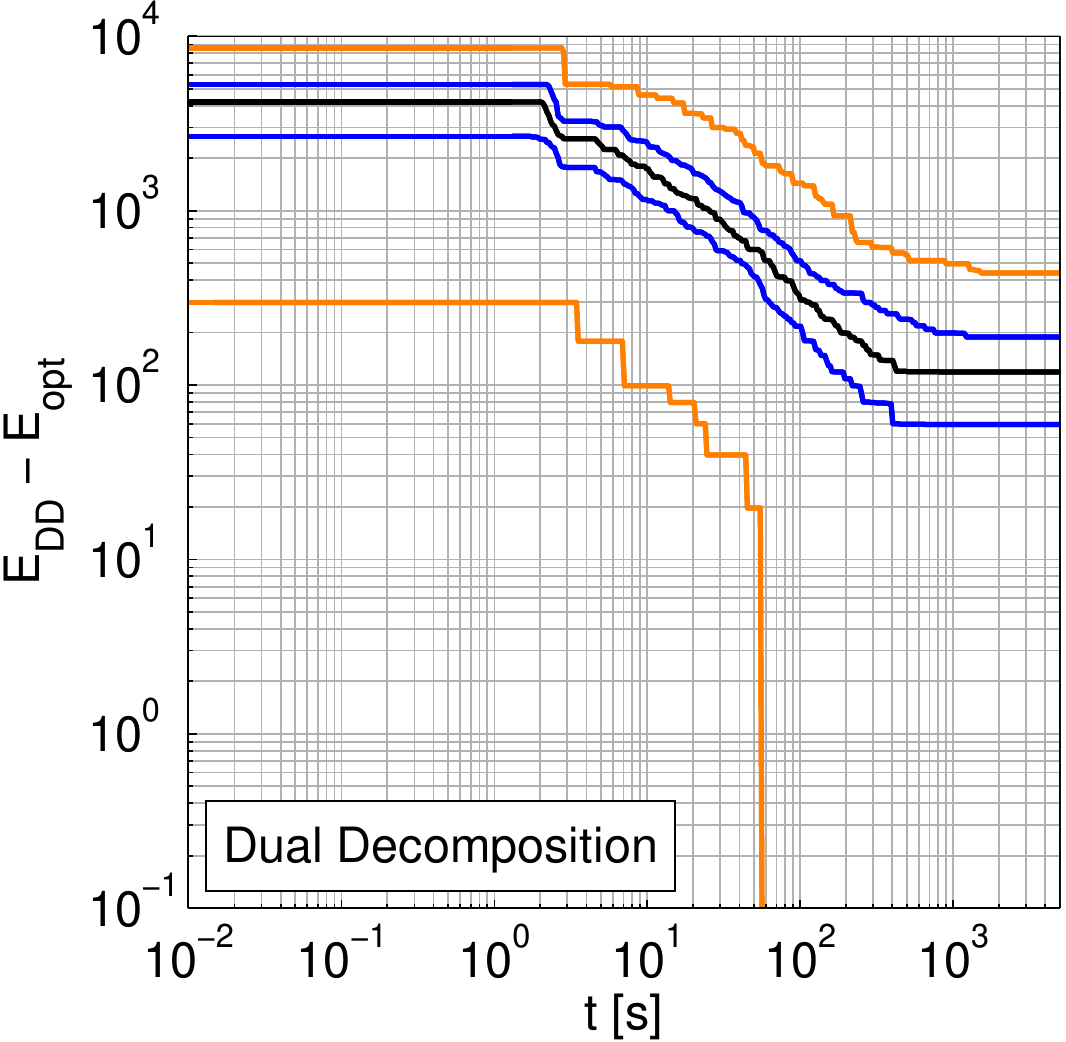} &
\includegraphics[height=5.6cm]{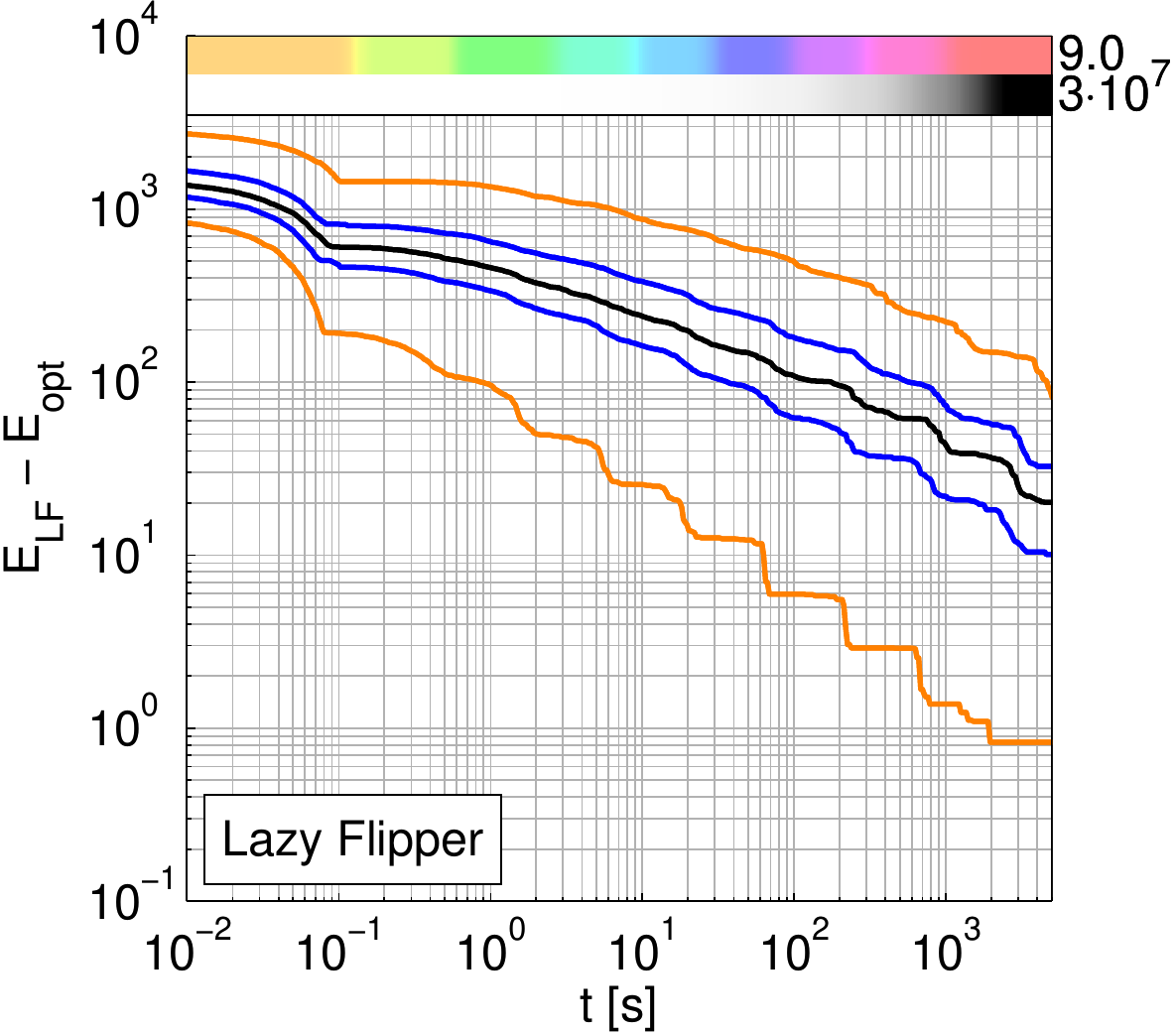}
\end{tabular}
\includegraphics[height=4.8cm]{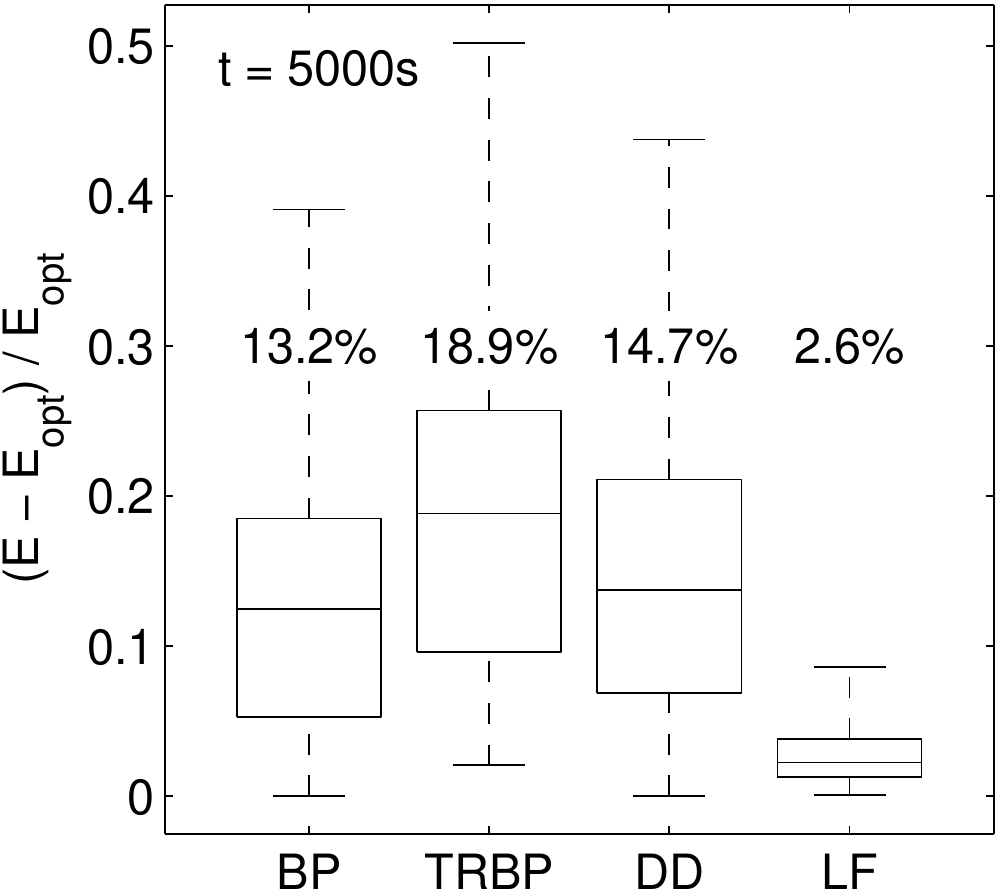}
\caption{Approximate solutions to the problem of removing excessive boundaries 
from over-segmentations of natural images. The search depth of the Lazy 
Flipper, averaged over all models in the ensemble, ranges from 1 (orange) to 9 
(red). At $t=5000$~s, $3 \cdot 10^7$ subsets are stored in the CS-tree.}
\label{figure:2d-seg-problem-curves}
\end{figure}

BP, TRBP, DD and the Lazy Flipper solve these problems approximately, thus 
providing upper bounds on the minimum energy. The differences between these 
bounds and the global optimum found by means of MILP are depicted in
Fig.~\ref{figure:2d-seg-problem-curves}.
It can be seen from this figure that, after 200 seconds, Lazy Flipper
approximations provide a tighter upper bound on the global minimum in the 
median than those of the other three algorithms. BP and DD have a better peak
performance, solving one problem to optimality. The Lazy Flipper reaches a 
search depth of 9 after around 1000 seconds for these sparse graphical models
using roughly 720~MB of RAM for the CS-tree. At $t=5000$~s and on average over
all models, its approximations deviate by only 2.6\% from the global optimum.

\subsection{Pruning of 3D Over-Segmentations}
\label{section:3d-segmentation-model}

The model described in the previous section is now applied in 3D to remove 
excessive boundaries from the over-segmentation of a volume image. In an 
ensemble of 16 such models obtained from 16 SBFSEM volume images, models have 
on average $16748 \pm 1521$ binary variables (and first order potentials), 
$26379 \pm 2502$ potentials of order~3, and $5081 \pm 482$ potentials of order~4.

\begin{figure}
\centering
\begin{tabular}{ll}
\includegraphics[height=5.6cm]{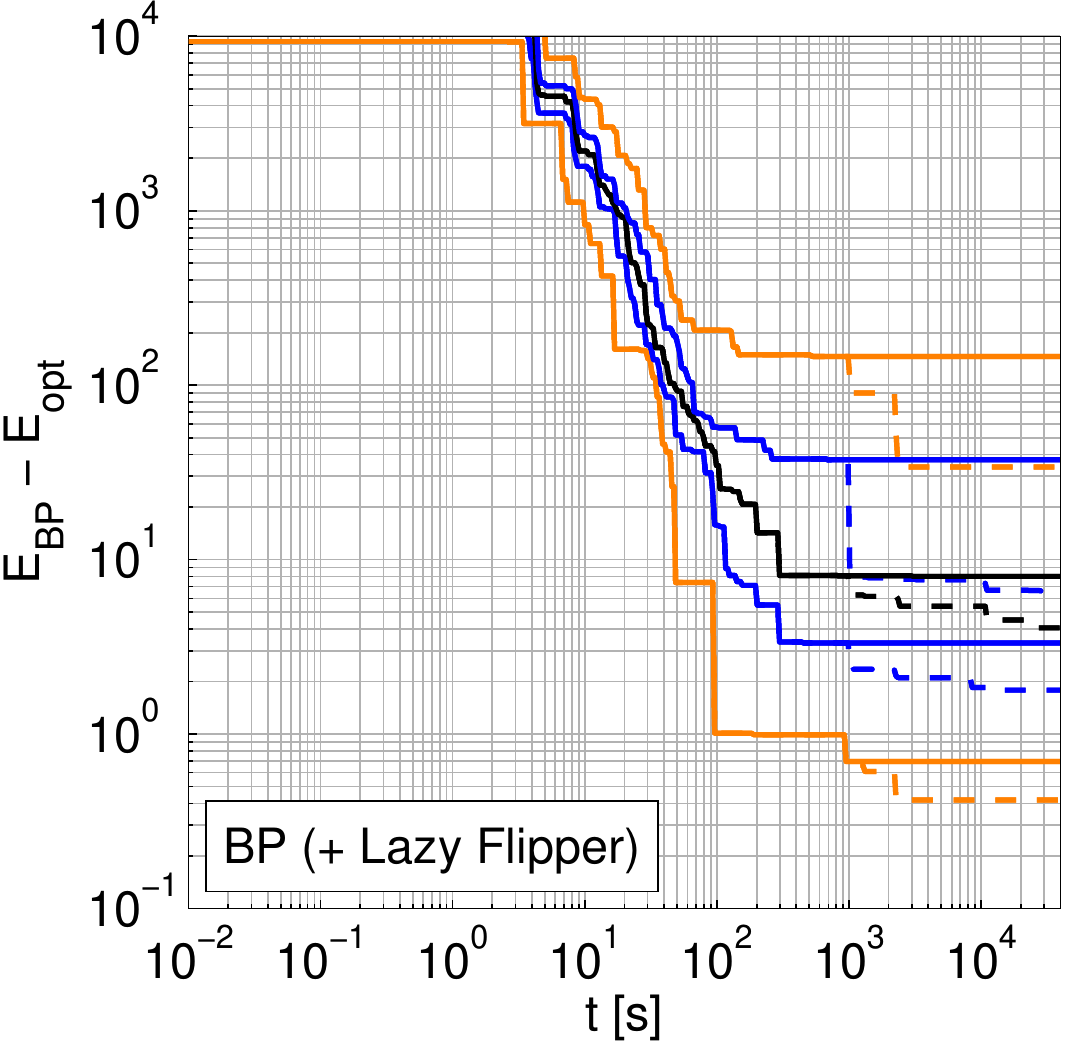} & 
\includegraphics[height=5.6cm]{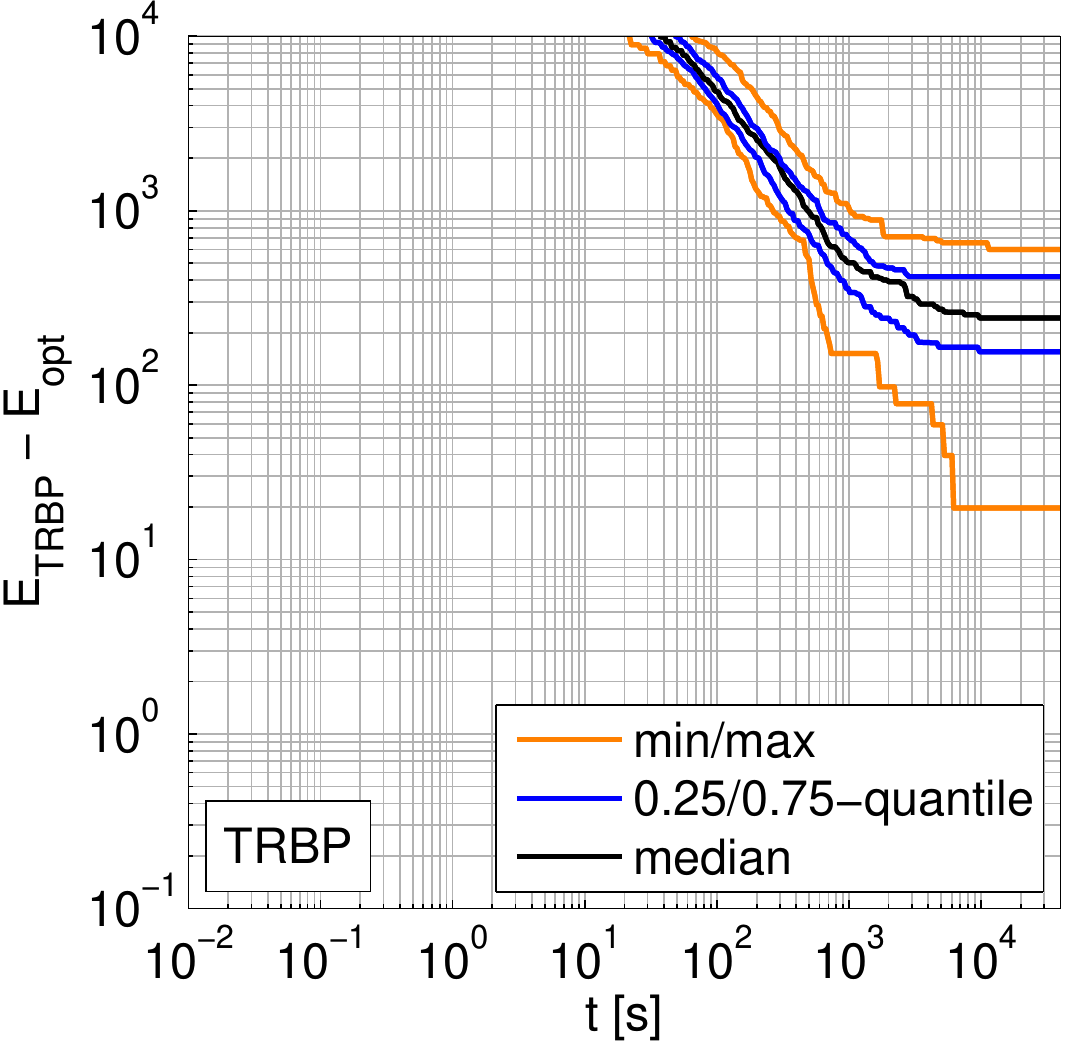} \\
\includegraphics[height=5.6cm]{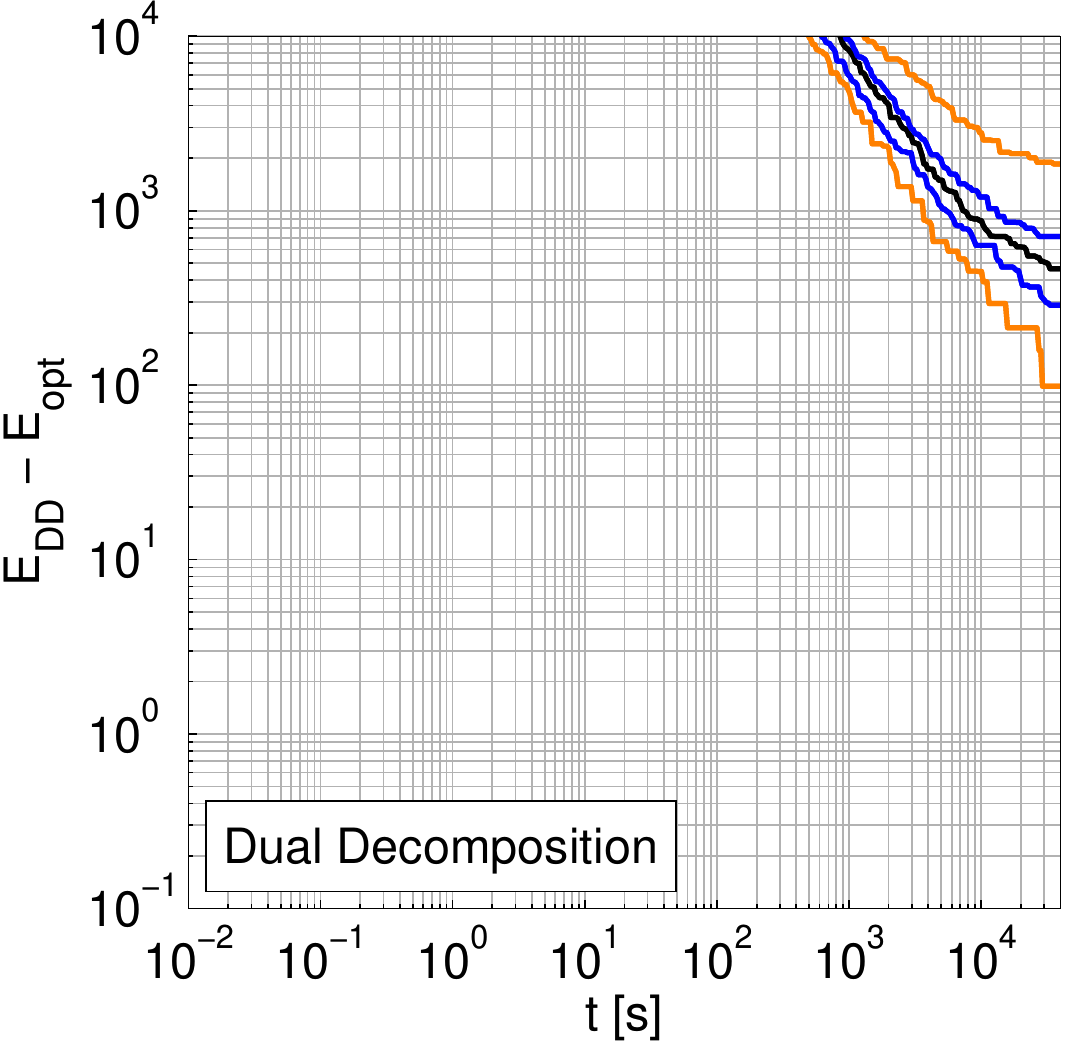} &
\includegraphics[height=5.6cm]{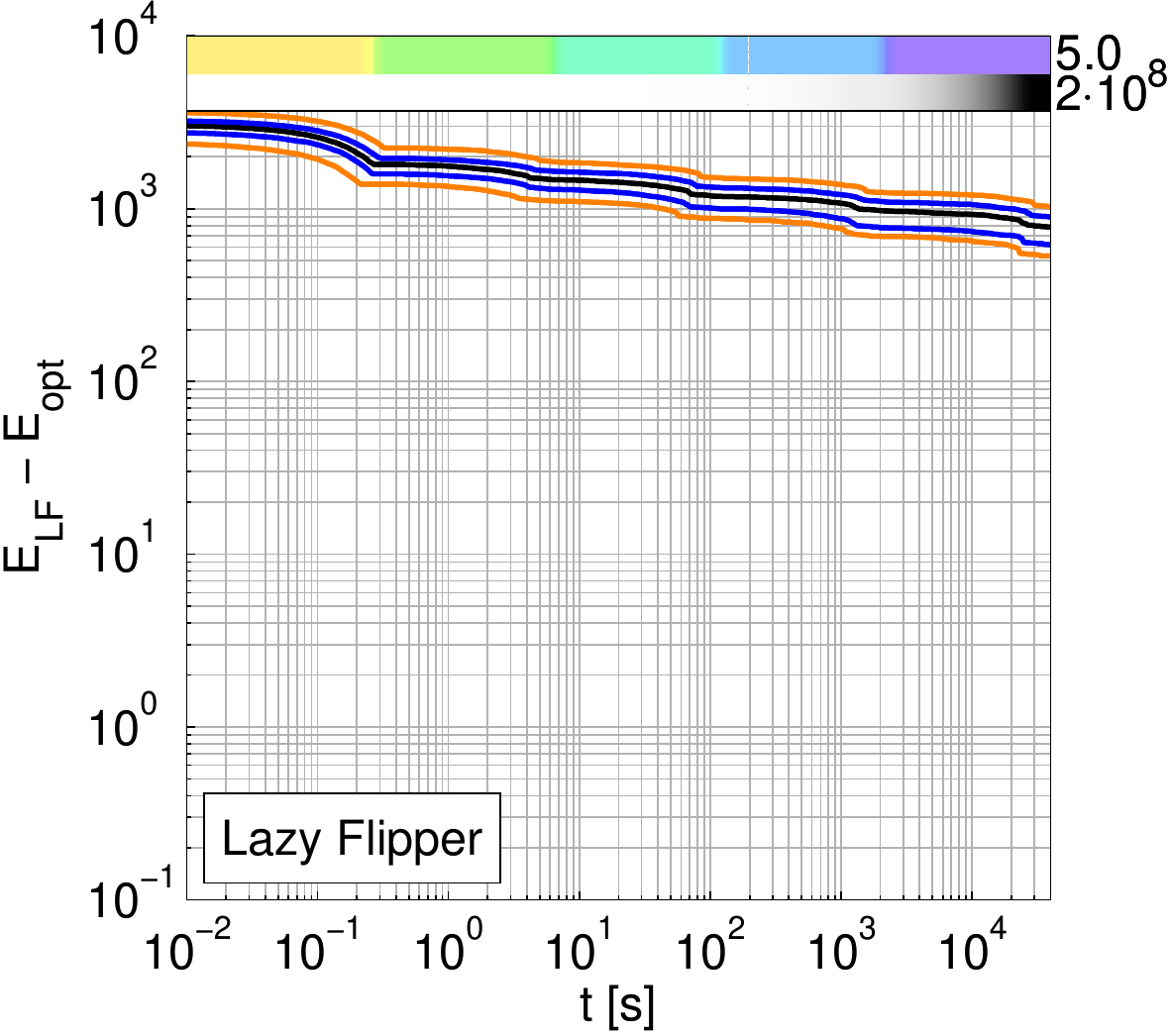}
\end{tabular}
\includegraphics[height=4.8cm]{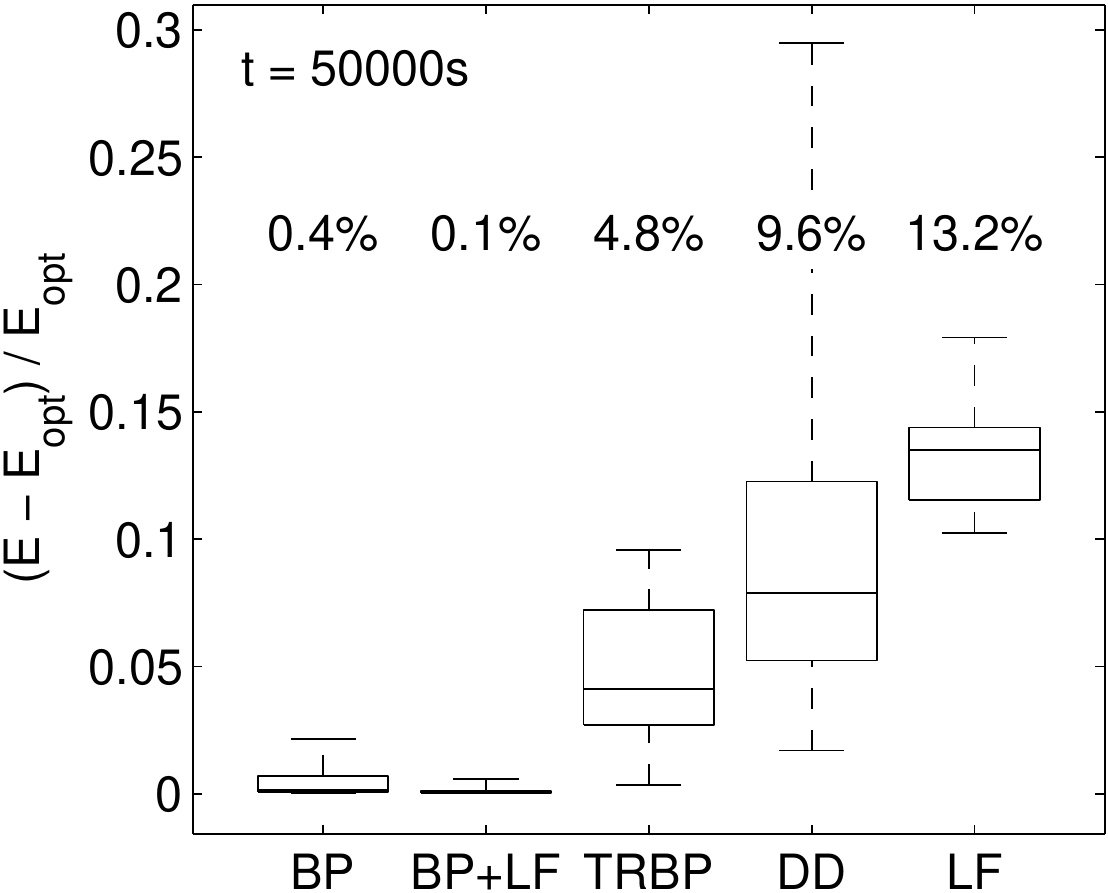}
\caption{Approximate solutions to the problem of removing excessive boundaries 
from over-segmentations of 3-dimensional volume images. The search depth of the 
Lazy Flipper, averaged over all models in the ensemble, ranges from 1 (yellow) 
to 5 (purple). After $50000$~s, $2 \cdot 10^8$ subsets are stored in the 
CS-tree. Dashed lines in the first plot show the result obtained when 
initializing the Lazy Flipper with the BP approximation at $t=100$s. This 
reduces the deviation from the global optimum at $t=50000$~s from 0.4\% 
on average over all models to 0.1\%.}
\label{figure:3d-seg-problem-curves}
\end{figure}

For BP, TRBP, DD and Lazy Flipper approximations, deviations from the global 
optimum are shown in 
Fig.~\ref{figure:3d-seg-problem-curves}.
It can be seen from this figure that BP performs exceptionally well on these
problems, providing approximations whose energies deviate by only 0.4\% on 
average from the global optimum. One reason is that most variables influence
many (up to 60) potential functions, and BP can propagate local evidence from 
all these potentials. Variables are connected via these potentials to as many as 
100 neighboring variables which hampers the exploration of the search space by 
the Lazy Flipper that reaches only of search depth of 5 after 10000 seconds, 
using 4.8~GB of RAM for the CS-tree, yielding worse approximations than BP, TRBP 
and DD for these models. 

In practical applications where volume images and the according models are 
several hundred times larger and can no longer be optimized exactly, it matters 
whether one can further improve upon the BP approximations. Dashed lines in the 
first plot in
Fig.~\ref{figure:3d-seg-problem-curves}
show the result obtained when initializing the Lazy Flipper with the BP 
approximation at $t=100$s. This reduces the deviation from the global optimum at 
$t=50000$~s from 0.4\% on average over all models to 0.1\%.

\section{Conclusion}

The optimum of a function of binary variables that decomposes according to a
graphical model can be found by an exhaustive search over only the connected 
subgraphs of the model. We implemented this search, using a CS-tree to 
efficiently and uniquely enumerate the subgraphs. 
The C++ source code is available from \url{http://hci.iwr.uni-heidelberg.de/software.php}. 
Our algorithm is guaranteed to converge to a global minimum when searching through
all subgraphs which is typically intractable. With limited runtime, 
approximations can be found by restricting the search to subgraphs of a given 
maximum size. Simulated and real-world problems exist for which these 
approximations compare favorably to those obtained by message passing and 
sub-gradient descent. For large scale problems, the applicability of the Lazy
Flipper is limited by the memory required for the CS-tree. However, for 
regular graphs, this limit can be overcome by an implicit representation of 
the CS-tree that is subject of future research.

\section*{Acknowledgments}

Acknowledgement pending approval by the acknowledged individuals.

\appendix
\section{Parameters and Model Decomposition}
\label{section:parameters}

In all experiments, the damping parameters for BP and TRBP are chosen optimally
from the set $\{0, 0.1, 0.2, \ldots, 0.9\}$. The step size of the sub-gradient 
descent is chosen according to
\begin{equation}
\tau_t = \alpha \frac{1}{1 + \beta t}
\label{eq:step-size-sequence}
\end{equation}
where $\beta = 0.01$ and $\alpha$ is chosen optimally from 
$\{0.01, 0.025, 0.05, 0.1, 0.25, 0.5\}$. The sequence of step sizes, in 
particular the function (\ref{eq:step-size-sequence}) and $\beta$ could 
be tuned further. Moreover, 
\cite{komodakis-2010}
consider the primal-dual gap and 
\cite{kappes-2010}
smooth the sub-gradient over iterations in order to suppress oscillations.
These measures can have substantial impact on the convergence.

The upper bounds obtained by BP, TRBP and DD do not decrease monoto\-nously. After 
each iteration of these algorithms, we therefore consider the elapsed runtime 
and the current best bound, i.e.~the best bound of the current and all preceding 
iterations. All five algorithms are implemented in C++, using the same optimized 
data structures for the graphical model and a visitor design pattern that allows 
us to measure runtime without significantly affecting performance. 

The same decomposition of each graphical model into tree models is used for 
TRBP and DD. Tree models are constructed in a greedy fashion, each comprising 
as many potential functions as possible. The procedure is generally applicable 
to irregular models with higher-order potentials: Initially, all potentials of 
the graphical model are put on a \emph{white list} that contains those 
potentials that have not been added to any tree model. A \emph{black list} of 
already added potentials and a \emph{gray list} of recently added potentials 
are initially empty. As long as there are potentials on the white list, new 
tree models are constructed. For each newly constructed tree model, the 
procedure iterates over the white list, adding potentials to the tree model if 
they do not introduce loops. Added potentials are moved from the white list to 
the gray list. After all potentials from the white list have been processed, 
potentials from the black list that do not introduce loops are added to the 
tree model. The gray list is then appended to the black list and cleared. The 
procedure finishes when the white list is empty. As recently shown in
\cite{kappes-2010},
decompositions into cyclic subproblems can lead to significantly tighter 
relaxations and better integer solutions.


\begin{thebibliography}{10}

\bibitem{cowell-2007}
Cowell, R.G., Dawid, A.P., Lauritzen, S.L., Spiegelhalter, D.J.:
\newblock Probabilistic Networks and Expert Systems: Exact Computational
  Methods for Bayesian Networks.
\newblock Springer (2007)

\bibitem{koller-2009}
Koller, D., Friedman, N.:
\newblock Probabilistic Graphical Models.
\newblock MIT Press (2009)

\bibitem{lauritzen-1996}
Lauritzen, S.L.:
\newblock Graphical Models.
\newblock Statistical Science. Oxford (1996)

\bibitem{wainwright-2008}
Wainwright, M.J., Jordan, M.I.:
\newblock Graphical Models, Exponential Families, and Variational Inference.
\newblock Now Publishers Inc., Hanover, MA, USA (2008)

\bibitem{besag-1986}
Besag, J.:
\newblock On the statisical analysis of dirty pictures.
\newblock Journal of the Royal Statistical Society B \textbf{48} (1986)
  259--302

\bibitem{boycov-2001}
Boykov, Y., Veksler, O., Zabih, R.:
\newblock Fast approximate energy minimization via graph cuts.
\newblock Transactions on Pattern Analysis and Machine Intelligence \textbf{23}
  (2001)  1222--1239

\bibitem{geman-1984}
Geman, S., Geman, D.:
\newblock Stochastic relaxation, gibbs distribution and the bayesian
  restoration of images.
\newblock Transactions on Pattern Analysis and Machine Intelligence \textbf{6}
  (1984)  721--741

\bibitem{mceliece-1998}
McEliece, R., MacKay, D., Cheng, J.F.:
\newblock Turbo decoding as an instance of pearl's belief propagation
  algorithm.
\newblock IEEE Journal on Selected Areas in Communications \textbf{16} (1998)
  140--152

\bibitem{pearl-1988}
Pearl, J.:
\newblock Probabilistic reasoning in intelligent systems: networks of plausible
  inference.
\newblock Morgan Kaufmann, San Francisco, CA, USA (1988)

\bibitem{kolmogorov-2004}
Kolmogorov, V., Zabin, R.:
\newblock What energy functions can be minimized via graph cuts?
\newblock Transactions on Pattern Analysis and Machine Intelligence \textbf{26}
  (2004)  147--159

\bibitem{schlesinger-2007}
Schlesinger, D.:
\newblock Exact solution of permuted submodular {MinSum} problems.
\newblock In: Proceedings of the 6th EMMCVPR. (2007)

\bibitem{schrijver-1986}
Schrijver, A.:
\newblock Theory of linear and integer programming.
\newblock John Wiley \& Sons, Inc., New York, NY, USA (1986)

\bibitem{schrijver-2003}
Schrijver, A.:
\newblock Combinatorial Optimization: Polyhedra and Efficiency.
\newblock Springer (2003)

\bibitem{kschischang-2001}
Kschischang, F.R., Frey, B.J., Loeliger, H.:
\newblock Factor graphs and the sum-product algorithm.
\newblock Transactions on Information Theory \textbf{47} (2001)  498--519

\bibitem{wainwright-2005}
Wainwright, M.J., Jaakkola, T., Willsky, A.S.:
\newblock {MAP} estimation via agreement on trees: message-passing and linear
  programming.
\newblock IEEE Transactions on Information Theory \textbf{51} (2005)
  3697--3717

\bibitem{komodakis-2010}
Komodakis, N., Paragios, N., Tziritas, G.:
\newblock {MRF} energy minimization and beyond via dual decomposition.
\newblock Transactions on Pattern Analysis and Machine Intelligence \textbf{99}
  (2010)

\bibitem{kappes-2010}
Kappes, J.H., Schmidt, S., Schnoerr, C.:
\newblock {MRF} inference by k-fan decomposition and tight {L}agrangian
  relaxation.
\newblock In: European Conference on Computer Vision 2010. (2010)

\bibitem{frey-2005}
Frey, B.J., Jojic, N.:
\newblock A comparison of algorithms for inference and learning in
  probabilistic graphical models.
\newblock Transactions on Pattern Analysis and Machine Intelligence \textbf{27}
  (2005)  1392--1416

\bibitem{jung-2009}
Jung, K., Kohli, P., Shah, D.:
\newblock Local rules for global {MAP}: When do they work?
\newblock In Bengio, Y., Schuurmans, D., Lafferty, J., Williams, C.K.I.,
  Culotta, A., eds.: Advances in Neural Information Processing Systems 22.
\newblock (2009)  871--879

\bibitem{swendsen-1987}
Swendsen, R.H., Wang, J.S.:
\newblock Nonuniversal critical dynamics in monte carlo simulations.
\newblock Physical Review Letters \textbf{58} (1987)  86--88

\bibitem{wolff-1989}
Wolff, U.:
\newblock Collective monte carlo updating for spin systems.
\newblock Physical Review Letters \textbf{62} (1989)  361--364

\bibitem{barbu-2003}
Barbu, A., Zhu, S.C.:
\newblock Graph partition by {Swendsen-Wang} cuts.
\newblock In: Proceedings of the Ninth IEEE International Conference on
  Computer Vision, Washington, DC, USA, IEEE Computer Society (2003)  320

\bibitem{minka-2001}
Minka, T.P.:
\newblock Expectation propagation for approximate {Bayesian} inference.
\newblock In: Proceedings of the 17th Conference on Uncertainty in Artificial
  Intelligence. (2001)  362--369

\bibitem{globerson-2007}
Globerson, A., Jaakkola, T.:
\newblock Fixing max-product: Convergent message passing algorithms for {MAP}
  {LP}-relaxations.
\newblock In: Advances in Neural Information Processing Systems, Cambridge, MA,
  USA, MIT Press (2007)  553--560

\bibitem{werner-2007}
Werner, T.:
\newblock A linear programming approach to max-sum problem: A review.
\newblock Transactions on Pattern Analysis and Machine Intelligence \textbf{29}
  (2007)  1165--1179

\bibitem{kohli-2008}
Kohli, P., Shekhovtsov, A., Rother, C., Kolmogorov, V., Torr, P.:
\newblock On partial optimality in multi-label {MRFs}.
\newblock In: Proceedings of the 25th International Conference on Machine
  Learning. (2008)

\bibitem{kumar-2009}
Kumar, M.P., Kolmogorov, V., Torr, P.H.S.:
\newblock An analysis of convex relaxations for {MAP} estimation of discrete
  {MRFs}.
\newblock Journal of Machine Learning Research \textbf{10} (2009)  71--106

\bibitem{kolmogorov-2006}
Kolmogorov, V.:
\newblock Convergent tree-reweighted message passing for energy minimization.
\newblock Transactions on Pattern Analysis and Machine Intelligence \textbf{28}
  (2006)  1568--1583

\bibitem{weiss-2001}
Weiss, Y., Freeman, W.:
\newblock On the optimality of solutions of the max-product belief-propagation
  algorithm in arbitrary graphs.
\newblock IEEE Transactions on Information Theory \textbf{47} (2001)  736 --744

\bibitem{moerkotte-2006}
Moerkotte, G., Neumann, T.:
\newblock Analysis of two existing and one new dynamic programming algorithm
  for the generation of optimal bushy join trees without cross products.
\newblock In: Proceedings of the 32nd International Conference on Very Large
  Data Bases. (2006)

\bibitem{zhang-2010}
Zhang, L., Ji, Q.:
\newblock Image segmentation with a unified graphical model.
\newblock Transactions on Pattern Analysis and Machine Intelligence \textbf{32}
  (2010)  1406--1425

\bibitem{dakin-1965}
Dakin, R.J.:
\newblock A tree-search algorithm for mixed integer programming problems.
\newblock Computer Journal \textbf{8} (1965)  250--255

\bibitem{land-1960}
Land, A.H., Doig, A.G.:
\newblock An automatic method of solving discrete programming problems.
\newblock Econometrica \textbf{28} (1960)  497--520

\bibitem{martin-2001}
Martin, D., Fowlkes, C., Tal, D., Malik, J.:
\newblock A database of human segmented natural images and its application to
  evaluating segmentation algorithms and measuring ecological statistics.
\newblock In: Proceedings of the 8th International Conference on Computer
  Vision (ICCV). Volume~2. (2001)  416--423

\bibitem{denk-2004}
Denk, W., Horstmann, H.:
\newblock Serial block-face scanning electron microscopy to reconstruct
  three-dimensional tissue nanostructure.
\newblock PLoS Biology \textbf{2} (2004)  e329

\bibitem{murphy-1999}
Murphy, K.P., Weiss, Y., Jordan, M.I.:
\newblock Loopy belief propagation for approximate inference: An empirical
  study.
\newblock In: Proceedings of Uncertainty in AI. (1999)  467--475

\end{thebibliography}
\end{document}